\documentclass[12pt,a4paper]{elsarticle}
\usepackage[table]{xcolor}
\usepackage{amsmath, amscd, amsthm, amsfonts, amssymb, graphicx, subcaption, color, soul, enumerate, xcolor,  mathrsfs, latexsym, bigstrut, framed, caption, mathtools, cancel, rotating}
\usepackage{pstricks,pst-node,pst-tree}
\usepackage{multirow}
\usepackage[pdftoolbar=true, bookmarksnumbered, colorlinks, plainpages, backref]{hyperref}
\AtBeginDocument{%
\hypersetup{linkcolor=darkblue, anchorcolor=green, citecolor=midblue, urlcolor=darkblue, filecolor=magenta}
}

\usepackage{nicematrix}
\usepackage{algorithm}
\usepackage{algpseudocode}
\def\NoNumber#1{{\def\alglinenumber##1{}\State #1}\addtocounter{ALG@line}{-1}}

\usepackage{etoolbox}

\usepackage{setspace}

\usepackage[normalem]{ulem}

\definecolor{cupgreen}{rgb}{0,0.498,0.208}
\definecolor{cupblue}{rgb}{0,0,.5}
\definecolor{cupred}{rgb}{1,0.04,0}
\definecolor{cuppink}{rgb}{0.925,0,0.545}
\definecolor{cupmagenta}{rgb}{0.624,0.161,0.424}
\definecolor{cupbrown}{rgb}{0.71,0.212,0.133}
\definecolor{vibtilg}{HTML}{00FF99}
\definecolor{C_ev}{HTML}{FF8C00}
\definecolor{C_evlmc}{HTML}{FF0000}
\definecolor{C_evls}{HTML}{802F00}
\definecolor{C_mm}{HTML}{4169E1}
\definecolor{C_mmlmc}{HTML}{228B22}
\definecolor{C_mmrls}{HTML}{4B0082}
\definecolor{des}{HTML}{edeb77}

\definecolor{TITLE}{rgb}{0,0,0}
\definecolor{midblue}{rgb}{0.00,0.0,0.80}
\definecolor{darkblue}{rgb}{0.00,0.00,0.45}
\definecolor{SECTION}{rgb}{0.50,0.00,1.00}
\definecolor{THM}{rgb}{0.8,0,0.1}
\definecolor{SEC}{rgb}{0,0,1}
\definecolor{calpolypomonagreen}{rgb}{0.12, 0.3, 0.17}
\definecolor{byzantine}{rgb}{0.64, 0.14, 0.54}
\definecolor{cadmiumgreen}{rgb}{0.0, 0.42, 0.24}
\definecolor{caribbeangreen}{rgb}{0.0, 0.6, 0.46}

\definecolor{col0}{RGB}{  0,   0,   0}%
\definecolor{col1}{RGB}{255,   0,   0}%
\definecolor{col2}{RGB}{  50, 156, 205}%
\definecolor{col3}{RGB}{  0, 128,   0}%
\definecolor{col4}{RGB}{  0,   0, 255}%
\definecolor{col5}{RGB}{255,   0, 255}%
\definecolor{col6}{RGB}{2, 224, 80}%
\definecolor{col7}{RGB}{165,  42,  42}%
\definecolor{col8}{RGB}{128, 128,   0}%
\definecolor{col9}{RGB}{255, 165,   0}%
\definecolor{col10}{RGB}{128,   0, 128}%
\definecolor{lightgray}{RGB}{176,176,176}

\makeatother
\normalsize

\theoremstyle{definition}

\numberwithin{equation}{section}
\newtheorem*{definition*}{\color{THM}Definition}
\newtheorem*{observation*}{\color{THM}Observation}

\journal{}

\begin{document}
	\begin{frontmatter}
		\title{Finding happiness by evolutionary algorithms}
		\author[]{Mohammad H. Shekarriz}
		\ead{m.shekarriz@deakin.edu.au}
		\affiliation{
			organization={School of Information Technology, Deakin University},
			city={Geelong},
			state={VIC},
			country={Australia}}
		\author[]{Dhananjay Thiruvady}
		\ead{dhananjay.thiruvady@deakin.edu.au}
		\author[]{Asef Nazari}
		\ead{asef.nazari@deakin.edu.au}

		\begin{abstract}
            A recent line of research concerns the problem of soft happy colouring (SHC), which requires that a partially coloured graph be extended to a complete colouring to maximise local agreements, so that as many vertices as possible end up surrounded by enough same-coloured neighbours. It is already known that SHC is NP-hard, and its solutions have a direct relationship with the community structure of networks; thus, it has immense applications in security and resilience. Past studies have shown that local search approaches can be fast and effective to an extent on the SHC; however, they often get stuck in local optima. Regarding the related problem of maximising happy vertices, evolutionary approaches have been proven effective; hence, this study develops a customised memetic algorithm that is a hybrid of genetic algorithms and local search. The experimental evaluation on a range of graphs in the stochastic block model shows that the memetic algorithm can achieve excellent results in search for an optimised solution to SHC compared to the local search approaches and standard genetic algorithms. Moreover, learning and evolution in the memetic algorithm allow diversification of solutions generated by fast, effective local search approaches, which prove superior for the challenging problem of community detection.
		\end{abstract}
		\begin{keyword}
			Soft happy colouring, metaheuristics, evolutionary algorithms, community detection, stochastic block model
		\end{keyword}
		
	\end{frontmatter}
	
	\section{Introduction}\label{Sec:Intro}

Partitioning network members into groups with shared interests or behavioural patterns remains a fundamental yet computationally demanding challenge in computer science. Recently, research has focused on \emph{Soft Happy Colouring} (SHC), a graph colouring extension introduced by Zhang and Li~\cite{ZHANG2015117} that maximises colour agreements in local neighbourhoods. Formally, SHC seeks to maximise the number of \emph{$\rho$-happy vertices} (for $0\leq \rho\leq 1$), defined as vertices where the proportion of same-coloured neighbours is at least $\rho$.

\begin{figure}
    \centering
    \includegraphics[scale=0.25]{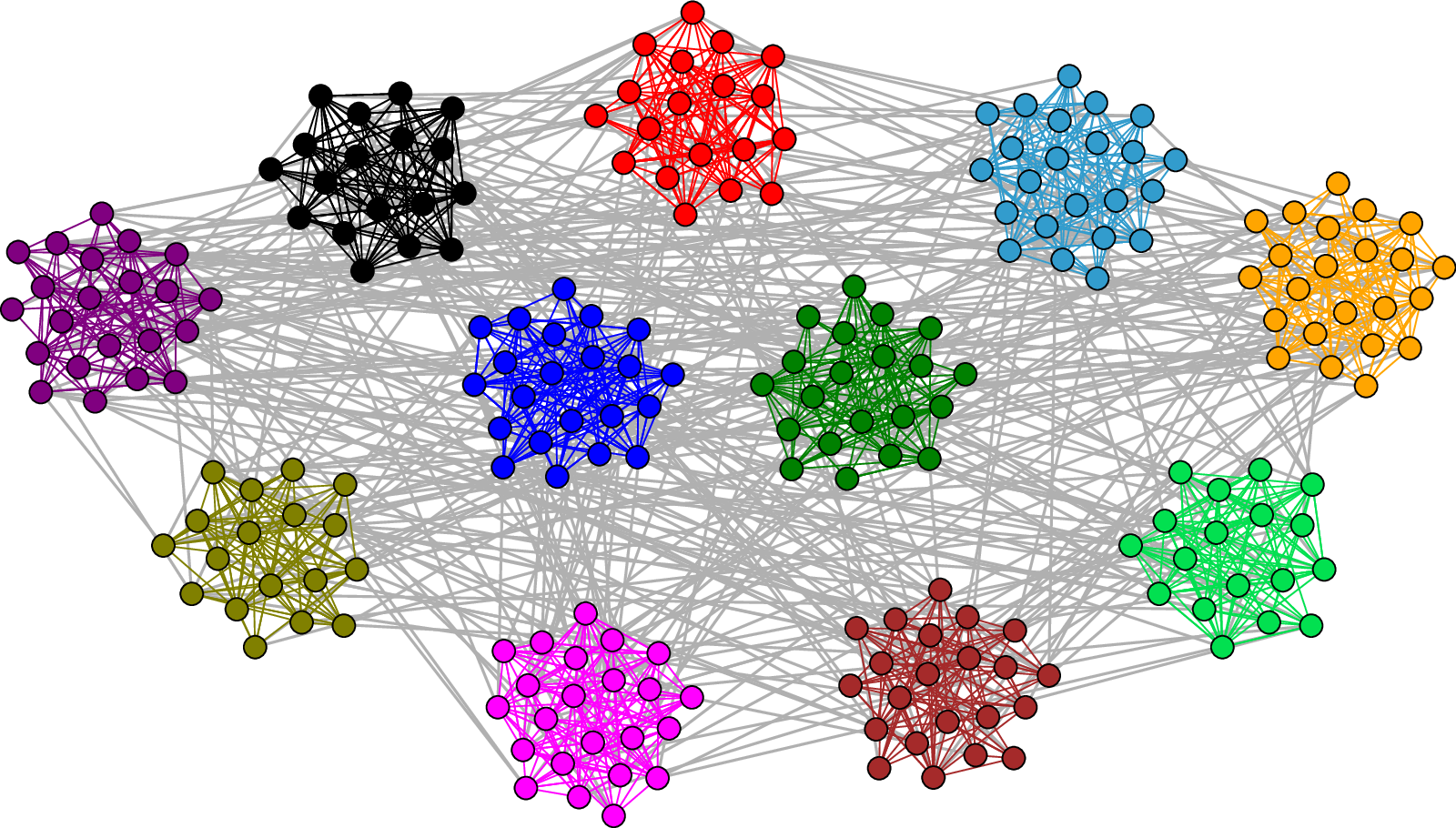}
    \caption{A connected graph $G$ with 11 communities. Grey lines are inter-community edges. The colouring $\sigma$ here is aligned with the communities and is a complete 0.6-happy colouring. The objective of \emph{community detection} is to find such a vertex partition.}\label{fig:communities}
\end{figure}

This formulation is grounded in the sociological principle of \emph{homophily}~\cite{Homophily}, which is effectively embodied by the community structure of graphs~\cite{Dev2016}. A \emph{community}~\cite{10.1007/978-3-540-48413-4_23} represents a cohesive group of vertices where internal connections are markedly denser than external ones (see Figure~\ref{fig:communities}).

Recent theoretical advances~\cite{SHEKARRIZ2025106893} demonstrate that communities can present a complete solution to SHC. Furthermore, Shekarriz et al.~\cite{SHEKARRIZ_local_search} verified that the colour classes of a complete SHC reveal underlying communities when $\rho$ falls within specific intervals. To tackle this, heuristic algorithms such as {\sf Local Maximal Colouring (LMC)}, {\sf Local Search (LS)}, and {\sf Repeated Local Search (RLS)} were introduced. While these methods are fast and effective at exploitation, they suffer from a critical limitation: they are prone to stagnation in local optima and lack the diversification necessary to explore the wider search space~\cite{SHEKARRIZ2025106893, SHEKARRIZ_local_search}.

This limitation highlights the need for more robust global search mechanisms. In the specific case where $\rho=1$ (known as "happy colouring"~\cite{ZHANG2015117}), finding a solution is notoriously difficult. Previous attempts have employed integer programming with CMSA~\cite{Lewis2019265} and tabu search~\cite{thiruvady2020,LEWIS2021105114}. Notably, Thiruvady et al.~\cite{THIRUVADY2022101188} demonstrated that evolutionary and hybrid approaches significantly outperform other methods in both diversification and intensification. This success provides a compelling motivation to engineer evolutionary algorithms for the more general and complex problem of SHC ($0 \leq \rho \leq 1$).

Evolutionary algorithms (EAs)~\cite{eiben2015introduction} offer an adaptive framework ideal for the optimisation challenges inherent in real-world networks~\cite{Goldberg1988}. By emulating natural selection mechanisms such as mutation and crossover, EAs can effectively navigate the large multimodal solution spaces found in network analysis~\cite{deb2011multi}. Their flexibility allows for balancing conflicting requirements of performance and robustness, making them suitable for dynamic environments~\cite{10.1093/oso/9780195099713.001.0001}.

However, standard \emph{Genetic Algorithms} (GAs), while robust in global search, often exhibit slow convergence due to stochastic exploration~\cite{chemin2024genetic}. To address this, we turn to \emph{Memetic Algorithms} (MAs), which hybridise evolutionary search with local optimisation. MAs have proven highly effective in overcoming the NP-hard nature of community detection~\cite{MORADI2019457}, allowing for the simultaneous enhancement of intra-community density while reducing inter-community links~\cite{Liu2019, MORADI2019457}. Given the proven local exploitation of heuristics like {\sf LS} and the global exploration of EAs, a hybrid memetic framework offers the ideal strategy for solving SHC.

In this paper, we develop and evaluate a specialised Memetic Algorithm framework for SHC. We investigate initialisation strategies using random or heuristic solutions (such as {\sf LMC} or {\sf LS}) and integrate rigorous local search algorithms into the evolutionary process. Our experiments on a large set of randomly generated graphs demonstrate that MAs provide statistically significant improvements over GAs, both in terms of SHC optimisation and community detection accuracy. Notably, the MA integrating {\sf RLS} over an {\sf LS}-generated population achieved the highest rank for SHC, while MAs using {\sf LMC} initialisation proved superior for community recovery. These results confirm that memetic approaches can effectively identify structures in real-world systems, from social networks to biological neural networks.

The remainder of this paper is organised as follows. Section~\ref{sec:nomen} establishes the technical background and algorithmic insights into SHC. Section~\ref{sec:method} details our evolutionary methodology, followed by the experimental analysis in Section~\ref{sec:experimental}. Finally, Section~\ref{sec:conc} concludes the paper and outlines directions for future research.

\section{Nomenclature}\label{sec:nomen} %

Networks are often modelled by graphs, and solutions to the optimisation problem of SHC are coloured graphs. A graph $G=(V(G), E(G))$ is a combinatorial object consisting of its vertex and edge sets~\cite{Chartrand-graphs_and_digraphs}. For $k\in\mathbb{N}$, a $k$-colouring is an onto function from the vertex set $V(G)$ to a set of $k$ colours. A partial $k$-colouring is a function defined on a proper subset of the vertex set onto a $k$-set of colours, the one that can be extended to a (complete) $k$-colouring.

\subsection{Communities}\label{sec:communities}

Real-world networks frequently exhibit community structures. In social networks, for instance, distinct groups often form around common interests or affiliations, thereby facilitating efficient information exchange and collective behaviour~\cite{PhysRevE.74.036104}. In the realm of cybersecurity, network architectures often reveal modular divisions that can mirror coordinated clusters of activity~\cite{Wang2021}. Similarly, neural networks display community structure through the formation of functionally specialised regions within the brain~\cite{10.7551/mitpress/8476.001.0001}. 

Detecting communities helps increase the robustness of networks through enhancing the intercommunity links~\cite{CUI2018185}. These cross-disciplinary examples underscore that community structure is a fundamental attribute of complex systems, and, consequently, reveal the possible applications of SHC.

For consistency with previous research~\cite{SHEKARRIZ2025106893, SHEKARRIZ_local_search}, our graphs are modelled by the Stochastic Block model (SBM)~\cite{HOLLAND1983109, JERRUM1998155}, %
a random graph model~\cite{Bollobas_2001} tailored for community-structured graphs. Specifically, we use a simplified version of the SBM, denoted by $\mathcal{G}(n,k,p,q)$, which is the probability space of $n$-vertex graphs with $k$ nonoverlapping (i.e. vertex-disjoint) communities. The edge probability inside a community is $p$ while an intercommunity edge has the probability of $q$. To have a meaningful and detectable community structure is guaranteed, we always assume that $q$ is much smaller than $p$. Moreover, for simplicity and comparability to other available results in the literature, we assume that the size of communities is almost equal to $\frac{n}{k}$.

\subsection{The SHC problem}\label{sec:Problem}

Assume that certain vertices of a graph $G$ are precoloured with $k$ colours (where $k \geq 2$), and let $\sigma$ be a $k$-colouring that extends this partial colouring. %
In this context, a vertex $v$ is \emph{$\rho$-happy} if at least $\lceil \rho \cdot \deg(v) \rceil$ of its neighbours are assigned the colour $\sigma(v)$. A vertex colouring $\sigma$ is defined as a \emph{soft happy colouring} (with $k$ colours) for $G$ if it maximises the number of $\rho$-happy vertices among all such $k$-colouring extensions of the partial colouring.

Used notations in this paper follow notations of~\cite{SHEKARRIZ_local_search, SHEKARRIZ2025106893}. A vertex is \emph{free} if it is not precoloured. When all vertices of $G$ are $\rho$-happy by $\sigma$, we write $\sigma\in H_\rho$. The notation $H_\rho (G,\sigma)$ or simply by $H_\rho (\sigma)$ represents the number of $\rho$-happy vertices of a colouring $\sigma$ of the graph $G$, $G\in H_\rho$ means that communities of $G$ induce a complete $\rho$-happy colouring, and for a vertex colouring $\sigma$, the \emph{accuracy of community detection of $\sigma$}, denoted by $ACD(\sigma)$, is the ratio of vertices whose colours align with their community membership. The \emph{ratio of $\rho$-happy vertices} in a colouring $\sigma$ is 
\begin{equation}
	\alpha(\sigma)=\frac{H_\rho (\sigma)}{n}.
\end{equation}

As previously noted, the idea of SHC lies in discovering the community structure of\linebreak graphs~\cite{SHEKARRIZ2025106893}. %
More importantly, colour classes of a $\rho$-happy colouring can represent the graph's communities~\cite{SHEKARRIZ_local_search}. For example, the colouring $\sigma$ of the graph $G$ in Figure~\ref{fig:communities} demonstrates this relation as it is a complete 0.6-happy colouring whose colour classes are aligned with the communities of the graph. Therefore, in this case, we have $\alpha(\sigma)=1$ and $ACD(\sigma)=1$.

To demonstrate that SHC is intimately connected to the community structure of graphs, it is shown in~\cite{SHEKARRIZ2025106893} that if $\rho$ is sufficiently small, every vertex in a graph $G$ in the SBM can become $\rho$-happy in the colouring induced by the communities of $G$. More precisely, suppose that $G\in\mathcal{G}(n,k,p,q)$, $n=|V(G)|$, $2\leq k$, $0<q<p<1$, $0<\rho\leq 1$, and $0<\varepsilon<1$. It is proved that if $\rho$ is less than a threshold, 
then with a high probability the communities of $G$ induce a $\rho$-happy colouring on $G$~\cite[Theorem 3.1]{SHEKARRIZ2025106893}. The threshold is
\begin{equation}\label{xi}
	\xi=\max\left\{\min\left\{\mathrm{ln}\left(\frac{\frac{k}{n}\mathrm{ln}(\varepsilon)+p e +(k-1)q}{p+(k-1)q}\right),\; \frac{p}{p+(k-1)q}\right\}, \; 0\right\}.
\end{equation}
It is also established that this favourable outcome becomes more pronounced as the number of vertices increases. For a graph $G\in\mathcal{G}(n,k,p,q)$, it is stated in~\cite[Theorem 3.3]{SHEKARRIZ2025106893} and~\cite[Theorem 2.1]{SHEKARRIZ_local_search} that the probability of communities induce a complete $\rho$-happy colouring approaches 1 when $n\rightarrow \infty$ and $ 0\leq \rho <\tilde{\xi}$ where
\begin{equation}\label{eq:xi_tilde}
	\tilde{\xi}=\lim_{n\rightarrow\infty} \xi=\frac{p}{p+(k-1)q}.  
\end{equation} 
However, when $\rho>\tilde{\xi}$, it is very unlikely to find a complete $\rho$-happy colouring, and when $n\to\infty$, almost no $\rho$-happy vertex would be found~\cite[Theorem 3.2]{SHEKARRIZ_local_search}.

When $\rho$ is very small, any colouring with no relation to community structure can have a large proportion of $\rho$-happy vertices. For a graph in the SBM $G\in\mathcal{G}(n,k,p,q)$, the lower bound
\begin{equation}\label{eq:mu}
	\mu=\frac{q}{p+(k-1)q}
\end{equation}
is stated for $\rho$ to expect a complete $\rho$-happy colouring has high correlation with the community structure of $G$~\cite{SHEKARRIZ_local_search}. When $0\leq \rho <\mu$, a vertex of one community can have the same colour as the vertices of other communities and still be $\rho$-happy. Consequently, realising the soft happiness condition for this case is easier, but the solution's colour classes might not give a good quality community detection~\cite{SHEKARRIZ_local_search}. 

For a graph in the SBM, the alignment with the original communities of the graph of a complete $\rho_2$-happy colouring is higher than that of a complete $\rho_1$-happy colouring when $\rho_1<\rho_2$~\cite[Theorem 3.1]{SHEKARRIZ_local_search}. %
And when $\mu\leq\rho\leq \tilde{\xi}$, colour classes of a complete $\rho$-happy colouring can present a community structure for a graph in the SBM~\cite{SHEKARRIZ_local_search}. %

\subsection{Known heuristics for SHC}

In~\cite{ZHANG2015117}, two heuristics were introduced for SHC. One of them, {\sf Greedy-SoftMHV}, gives only one colour to all the free vertices, the one with maximum $\rho$-happiness. The other, {\sf Growth-SoftMHV}, splits the free vertices into some groups, and prioritises the search of vertices of one group over another. {\sf NGC} is an extension to {\sf Greedy-SoftMHV}, which is introduced in~\cite{SHEKARRIZ2025106893}. We use neither of these algorithms for designing evolutionary algorithms because of their high time complexity, rigidity or independence from the community structure of the graph~\cite{SHEKARRIZ_local_search}. 

To design effective and reliable MAs, we need fast and effective heuristics with non-determin-\linebreak istic outputs. Slow heuristics such as {\sf Growth-SoftMHV} consume most of the time allocated to the MA to generate or enhance a population of solutions, while running a non-randomised heuristic such as {\sf Greedy-SoftMHV} many times gives the population score no enhancement privilege.

{\sf Local Maximal Colouring (LMC)} is introduced in~\cite{SHEKARRIZ2025106893} and has linear time complexity ($\mathcal{O}(m)$, where $m$ is the number of edges). Moreover, its output is highly correlated with the graph's community structure. In the design of the {\sf LMC}, there is no dependence on the proportion of happiness $\rho$, but some randomness is employed. In its main loop, the algorithm randomly chooses a vertex $v$ from the intersection of uncoloured vertices and neighbours of already coloured vertices, and colours $v$ with the most frequent colour in $N(v)$. 

\begin{figure}
	\centering
    \begin{tabular}{ c c c }
    \includegraphics[scale=0.28]{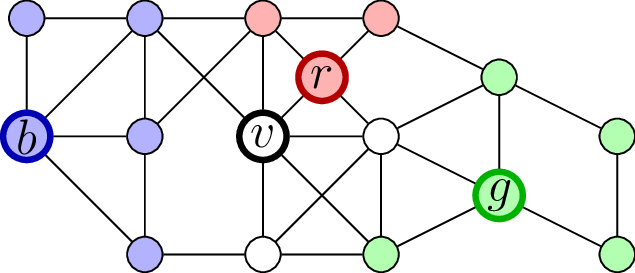}
    &
    \hspace{5mm} $\vcenter{\hbox{$\xrightarrow[]{\text{\sf LMC}}$}}$ \hspace{5mm}
    &
    \includegraphics[scale=0.28]{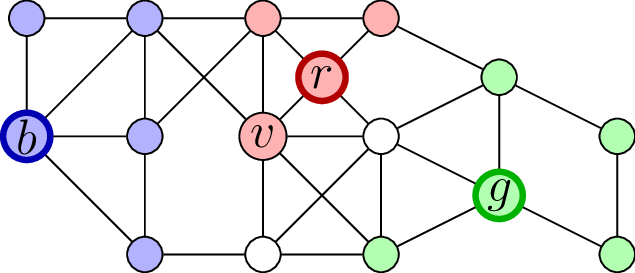}\\
    (a)& & (b)
    \end{tabular}
	\caption{The graph depicted on the left (a) is being coloured by {\sf LMC}. Vertices denoted by $r$, $g$ and $b$ are precoloured vertices. At this stage, the vertex $v$ is selected to be coloured. Because it has two red, one blue, one green, and one uncoloured neighbours, the colour red is given to it (b).}
	\label{fig:LMC}
\end{figure}

Figure~\ref{fig:LMC} illustrates {\sf LMC} for a small graph. Three vertices denoted by $r$, $g$ and $b$ are precoloured vertices whose colours, respectively red, green and blue, must not be changed. At a time, the vertex $v$ is selected. Because it has two red, one green and one blue neighbours (as well as two uncoloured neighbours), the colour red has the maximum number. Therefore, the colour red is given to $v$. 

Another linear-time ($\mathcal{O}(m)$) algorithm for SHC is {\sf Local Search (LS)}, %
which is not only a heuristic but also an improvement algorithm. It is introduced in~\cite{SHEKARRIZ_local_search}, and runs as follows. It copies the input vertex colouring $\sigma$ as $\tilde{\sigma}$ and gathers all its $\rho$-unhappy free vertices in a set $U$. Then it examines all vertices $v\in U$, one by one in a random order, to see if $\tilde{\sigma}(v)$ agrees with the most frequent colour $q$ in $N(v)$. And if not, $\tilde{\sigma}(v)$ is changed to $q$.

\begin{figure}
	\centering
    \begin{tabular}{ c c c }
    \includegraphics[scale=0.28]{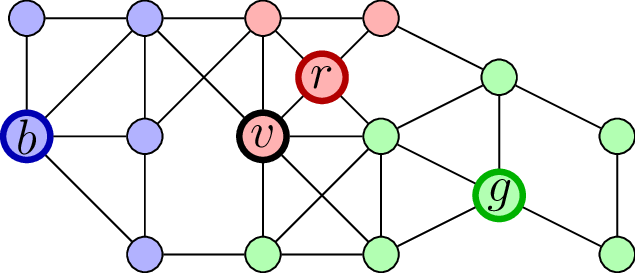}
    &
    \hspace{5mm} $\vcenter{\hbox{$\xrightarrow[]{\text{\sf LS}}$}}$ \hspace{5mm}
    &
    \includegraphics[scale=0.28]{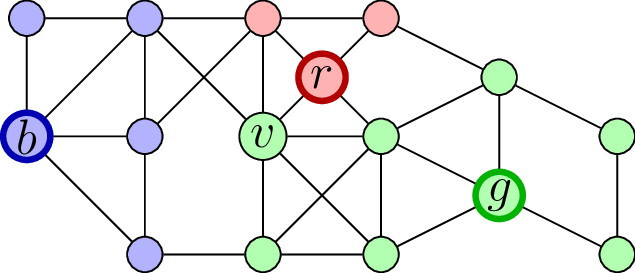}\\
    (a)& & (b)
    
    \end{tabular}
	\caption{The number of $\rho$-happy vertices in the coloured graph of (a) is getting improved by {\sf LS}, and the free vertex $v$ is selected. Because it has three green, two red, and one blue neighbours, its colour is changed from red to green (b).}
	\label{fig:LS}
\end{figure}

The algorithm {\sf LS} is demonstrated in Figure~\ref{fig:LS}. A $\rho$-unhappy free vertex $v$ is selected to be checked. In its neighbours, there are three green vertices, two red and one blue. Its colour is red, but the most frequent colour in $N(v)$ is green. Therefore, its colour is changed to green.

Two other versions of local search algorithms were also introduced in~\cite{SHEKARRIZ_local_search}, namely {\sf Repeated Local Search (RLS)} and {\sf Enhanced Local Search (ELS)}. The former is similar to {\sf LS}, only refills the set $U$ with $\rho$-unhappy vertices when checking vertices in $U$ ends; however, the latter is a local search algorithm that checks the objective function, i.e. the number of $\rho$-happy vertices in the entire vertex set, any time it decides about a vertex colour. 

Due to a higher time requirement, {\sf ELS} produces poor results in a reasonable amount of time, and thus is not suitable for designing metaheuristics. %
On the other hand, although {\sf RLS} can be much more time costly if its input is a partially coloured or a randomly coloured graph, it is useful if it inputs a solution by {\sf Growth-SoftMHV} (which is very time costly on its own), or when it is run on a solution by {\sf LS}. In the latter case, it can make a slight difference in a reasonable amount of time~\cite{SHEKARRIZ_local_search}. As a result, it can be useful for designing metaheuristic algorithms for SHC, especially when a solution by {\sf LS} needs to be enhanced.  

The results of the practical tests show that {\sf LMC} and {\sf LS} are very fast and effective heuristics whose randomness in selecting $\rho$-unhappy free vertices changes their output every time they are used~\cite{SHEKARRIZ2025106893, SHEKARRIZ_local_search}. Thus, it is natural to use them in the design of evolutionary algorithms, especially to create initial populations. Furthermore, {\sf LS} is a fast and effective improvement algorithm, so it is an obvious choice for the enhancement needed in MAs~\cite{SHEKARRIZ_local_search}. Other improvement algorithms require significantly more time than {\sf LS}, and their use instead of {\sf LS} is not justified unless the initial population is made solely by {\sf LS}. When {\sf LS} is being used to form the initial population, it adds no enhancements if being run again, the case for which we use {\sf RLS} in our MAs.

\section{Hybridising Evolutionary Algorithms}\label{sec:method}

Evolutionary algorithms are a subset of metaheuristic optimisation techniques inspired by the principles of natural selection and biological evolution. They are widely used to solve complex optimisation problems, particularly when traditional methods struggle due to non-linearity, a vast or multi-dimensional search space, or the high complexity of other existing methods. By mimicking the evolutionary processes found in nature, namely reproduction, mutation, and selection, EAs aim to evolve solutions toward optimal or near-optimal configurations over iterative cycles, which are called generations. Due to the complex nature of SHC problems, it can be expected that EAs perform well on the SHC problem, and have in fact shown to be in \cite{THIRUVADY2022101188}.

\begin{figure}
	\centering
    \includegraphics[scale=0.27]{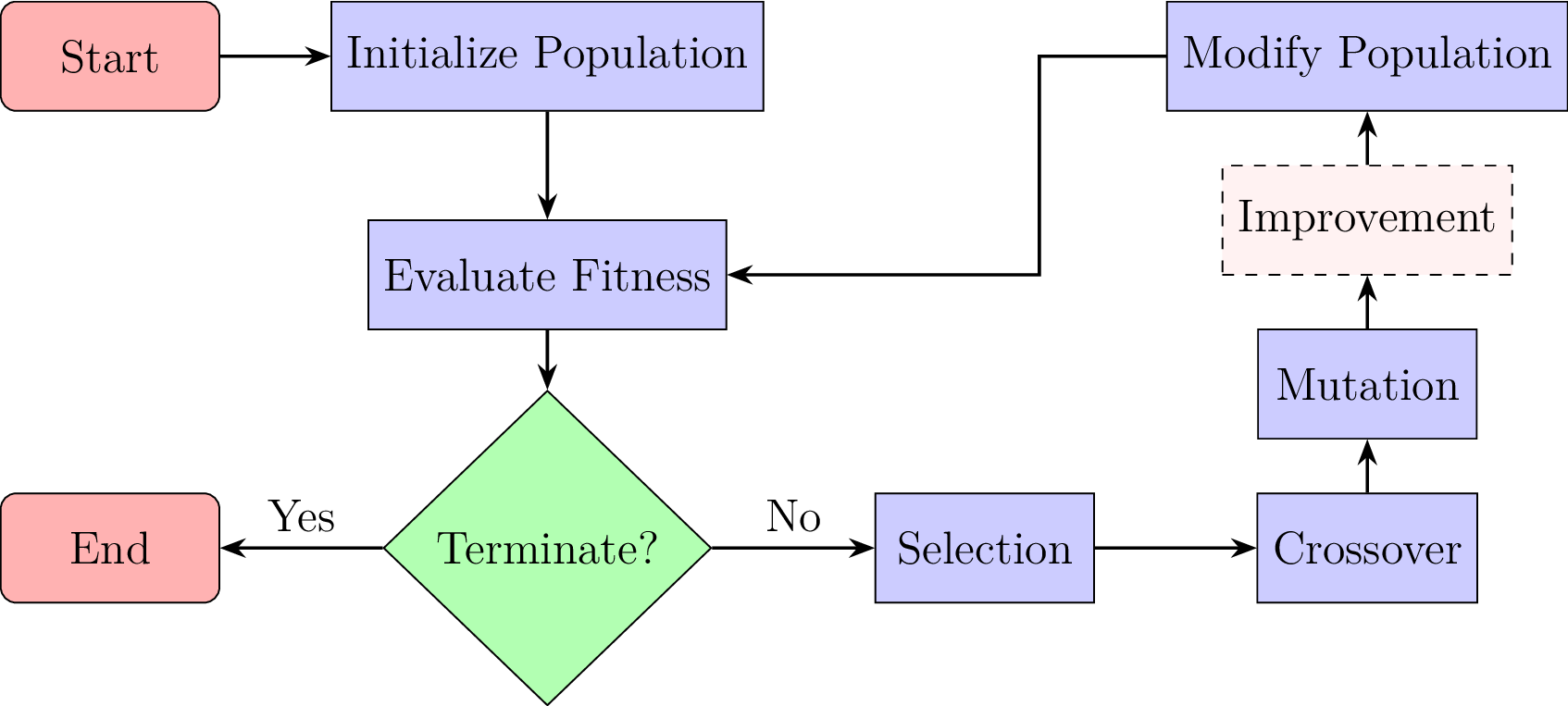}
	\caption{This is a flowchart of EAs. The improvement step (pale pink dashed-rectangle) is essential in MAs, but it is skipped in GAs.}
	\label{fig:ev-flowchart}
\end{figure}

The working mechanism of evolutionary algorithms begins with the initialisation of a population comprising feasible solutions. The initial populations of some EAs consist of totally random solutions, but having an initial population of feasible solutions with non-random structures %
could be a game-changer. %
At each generation, individual solutions from the population undergo evaluation through \emph{selection procedure} or a \emph{fitness function}, which in our case is selecting colourings with the highest numbers (or ratios) of $\rho$-happy vertices. %
Operators of EAs, namely crossover (recombination) and mutation, introduce diversity by combining the genetic material of parent solutions and institute small random alterations, respectively. This stochastic variation ensures exploration of large parts of the solution space and helps to avoid premature convergence. Figure~\ref{fig:ev-flowchart} presents a flowchart of evolutionary algorithms.%

Key operations like crossover and mutation play pivotal roles in evolutionary algorithms. %
Crossover combines segments of genetic information from parent individuals to produce offspring that inherit traits from both, fostering the exploitation of existing high-quality solutions. In contrast, mutation introduces random changes to an individual's genetic code, promoting exploration and maintaining diversity within the population. This balance between exploration and exploitation is essential for the efficacy of evolutionary algorithms, as it ensures both the thorough exploration of the search space and the refinement of promising areas. %

EAs can have another step, usually called improvement, shown in pink in Figure~\ref{fig:ev-flowchart}. MAs, as an advanced class of EAs, incorporate an improvement algorithm (usually a local search method) to enhance the optimisation process. They combine the global exploration capabilities of evolutionary strategies with the fine-tuning strength of local search techniques~\cite{eiben2015introduction}. %

The foundational concept of MAs draws inspiration from the idea of cultural evolution, where \emph{memes}, which are units of cultural information, spread and adapt through a population. Within MAs, individuals undergo not only genetic operations, such as crossover and mutation, but also a local optimisation phase that improves their fitness before proceeding to the next generation. This dual mechanism effectively speeds up its convergence, making it particularly adept at solving complex, large-scale optimisation problems. %
	It was previously shown that effective local search strategies can be devised for SHC~\cite{SHEKARRIZ_local_search}, thereby warranting memetic approaches for tackling this problem.%

We propose MAs for SHC and compare our approach to GAs. Each of these comes from three different types: (a) randomly generated initial population denoted by {\sf Rnd}, (b) initial populations resulting from the outputs of {\sf LMC}, and (c) initial populations resulting from the outputs of {\sf LS}. In total, we have six algorithms. The GA %
variants are {\sf GA(Rnd)}, {\sf GA(LMC)}, and {\sf GA(LS)} and the MA ones are {\sf MA(Rnd)}, {\sf MA(LMC)} and {\sf MA+RLS(LS)}, where the algorithms that produce the initial population are mentioned in parentheses. As mentioned, there was no point in improving a population generated by {\sf LS} by the same algorithm. Therefore, in this case, we use {\sf RLS} for the improvement step of the algorithm {\sf MA+RLS(LS)}. We tested these algorithms for a large set of randomly generated partially coloured graphs. The experimental results of these tests are reported in Section~\ref{sec:experimental}. 

There are several parameters to consider when devising the EAs, namely \emph{population size}, \emph{time limit}, and \emph{mutation factor}. A note on how these parameter settings are chosen is presented in Subsection~\ref{sec:test-dtls}. The EAs' inputs and outputs are as follows:
\begin{quote}
	\textbf{Input:} a partially $k$-coloured connected graph $G$, the proportion of happiness $\rho$, the population sizes $PopSize$, the mutation factor $MuteFactor$, and the set of precoloured vertices $V'$
	
	\textbf{Output:} a $k$-colouring $\tilde{\sigma}$ for $G$
\end{quote}

The main characteristic of GAs %
is that instead of improving one solution at a time, they work on a population of solutions. Hence, in Line 1 of Algorithm~\ref{alg:ev}, the initial solutions are generated using a fast heuristic. Therefore, the three variations of GAs are only different in their initial populations. 

\begin{algorithm}
	\caption{{\sf Genetic Algorithm (GA)} -- Searching for the optimal solution of $\rho$-happy colouring by combining and selecting solutions.}\label{alg:ev}
	
	\begin{algorithmic}[1]
		\State $P \gets$ a population of heuristic solutions of size $PopSize$
		\State $Scores\gets \{H_\rho (t), \forall t\in P\}$
		\State $\tilde{\sigma}\gets t$ for $t\in P$ such that $H_\rho (t)=\max Score$
		\NoNumber{ }
		\While{$TerminateCondition \neq$ True}
		\State $Parents\gets SelectParents(P,Scores)$
		\State $NewGen\gets CrossOver(Parents,PopSize, p)$\Comment{Usually $p=\frac{1}{2}$}
		\State $Mutate(NewGen, MuteFactor)$
		\State $P\gets Parents \cup NewGen$
		\State $Scores\gets \{H_\rho (t), \forall t\in P\}$
		\If{$\max Score> H_\rho (\tilde{\sigma})$}
		\State $\tilde{\sigma}\gets t$ for $t\in P$ such that $H_\rho (t)=\max Score$
		\EndIf
		\EndWhile
		\State {\bf Return} $\tilde{\sigma}$
	\end{algorithmic}
\end{algorithm}

After the initial population is generated, in Line 2, the fitness of each individual in the population is stored in a vector named $Scores$, and in Line 3, the \emph{best} colouring in the population is stored in $\tilde{\sigma}$. Then the main loop of the algorithm, Lines 4 to 13, is executed. The $TerminateCondition$ is usually whether the $TimeLimit$ has already passed or the number of $\rho$-happy vertices of $\tilde{\sigma}$ has already reached the maximum possible (if no maximum is known, it can be set to the number of vertices $n$). 

\begin{algorithm}
	\caption{{\sf CrossOver} -- Combining two solutions to generate a new solution called \emph{offspring}.}\label{alg:CrossOver}
	\begin{flushleft} $\;$\hspace*{\algorithmicindent}
		\textbf{Input:} $Parents$, $PopSize$, $p$ %
		\Comment{$p=$ probability of choosing a parent, usually $\frac{1}{2}$}\\
		\hspace*{\algorithmicindent} \textbf{Output:} $NewGen$ \Comment{$NewGen$ is a set of new solutions}
	\end{flushleft}
	\begin{algorithmic}[1]
		\State $NewGen \gets \emptyset$
		\NoNumber{}
		\While{$|NewGen|<PopSize - |Parents|$}
		\State {\bf Randomly Select} $\sigma_1, \sigma_2 \in Parents$
		\For{$v\in V\setminus V'$} \Comment{$V' = $ the set of precoloured vertices}
		\State $\sigma(v)\gets Bernolli\left(\{\sigma_1(v), \sigma_2(v)\},p\right)$
		\EndFor
		\State {$NewGen\gets\{\sigma\}\cup NewGen$}
		\EndWhile
		\State {\bf Return} $NewGen$
	\end{algorithmic}
\end{algorithm}

\begin{figure}
	\centering
    \begin{tabular}{ c c c c c c }
    & Parent 1 & & Parent 2 & & Offspring\\
    $\vcenter{\hbox{0)}}$& \includegraphics[scale=0.24]{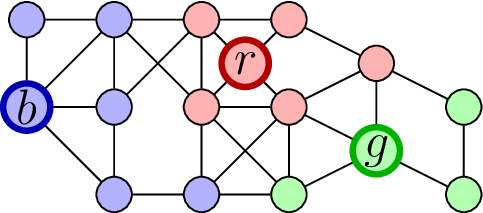}& &
    \includegraphics[scale=0.24]{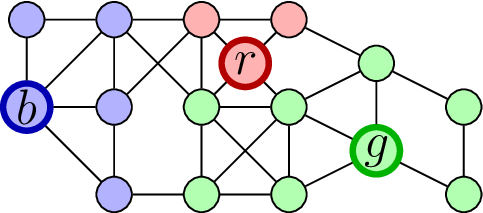}
    &
     \hspace{5mm} 
    &
    \includegraphics[scale=0.24]{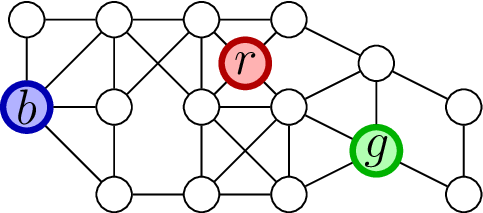}\\[1cm]

    $\vcenter{\hbox{1)}}$ &
    \includegraphics[scale=0.24]{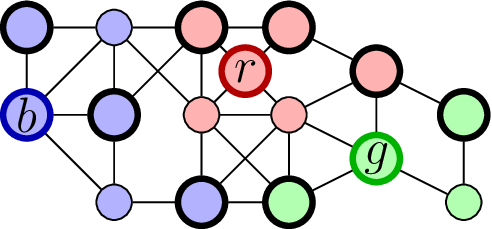}
    &
    &
    \includegraphics[scale=0.24]{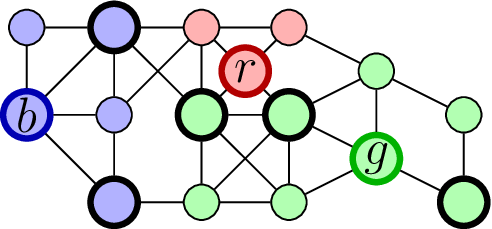}
    &
     $\vcenter{\hbox{ $\xrightarrow[]{\text{\sf CrossOver}}$ }}$
    &
    \includegraphics[scale=0.24]{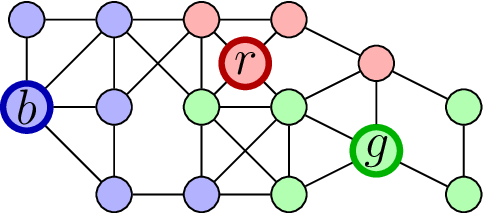}\\

    \end{tabular}
	\caption{The procedure {\sf CrossOver} starts by selecting two solutions  (parents). At the beginning (stage 0), the free vertices of the offspring are uncoloured. During {\sf CrossOver} (stage 1), some free vertices are randomly selected from the first parent and other free vertices are selected from the second parent. Then the colours of the selected vertices are transferred to the free vertices of the offspring.}
	\label{fig:CrossOver}
\end{figure}

$SelectParents$ in Line 5 chooses the most fit individuals in the population. In our tests, this procedure chooses the fittest half of the individuals in the population and stores these colourings in the vector $Parents$. Then, in Line 6, the procedure $CrossOver$ transforms the vector $Parents$ to a population of size $PopSize$. The procedure $CrossOver$ is given in Algorithm~\ref{alg:CrossOver} and illustrated in Figure~\ref{fig:CrossOver}. For each pair of parents, the procedure generates two offspring solutions to extend the set of $Parents$ to a complete population. For a vertex $v$ of $G$, with the probability $p=\frac{1}{2}$, one parent, say $\sigma_1$, is chosen and the colour $\sigma_1 (v)$ is given to the same vertex in an offspring. All offspring solutions are generated in a new vector $NewGen$ (of size $PopSize - |Parents|$).

\begin{figure}
	\centering
	\begin{tabular}{ c c c c c }

    \includegraphics[scale=0.24]{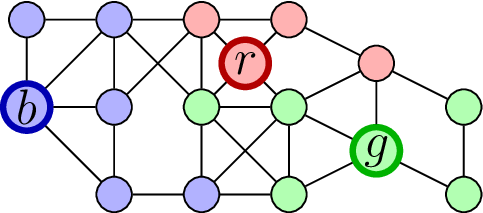}
    &
     $\vcenter{\hbox{ $\xrightarrow[]{\text{\sf Select}}$ }}$
    &
    \includegraphics[scale=0.24]{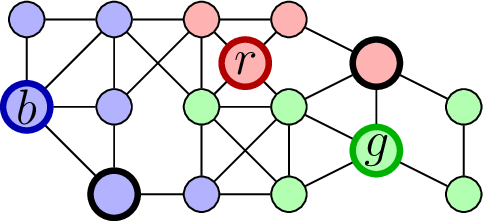}
    &
     $\vcenter{\hbox{ $\xrightarrow[]{\text{\sf Mutate}}$ }}$
    &
    \includegraphics[scale=0.24]{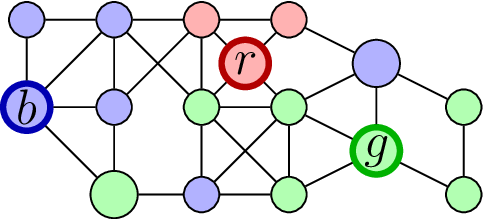}\\

    \end{tabular}
	\caption{For mutating, some free vertices are randomly selected, and their colours are randomly changed.}
	\label{fig:Mute}
\end{figure}

Mutation of the offspring happens in Line 7 of the algorithm by the function $Mutate$. If the mutation factor times the number of free vertices is denoted by $\mathfrak{m}$, for each offspring, $\mathfrak{m}$ random vertices from free vertices are chosen and their colours are randomly changed to colours from $1, \ldots, k$. Figure~\ref{fig:Mute} demonstrates this function for a small graph.%

Following this in Line 8, the population $P$ is formed again from the union of $NewGen$ and $Parents$. In the next line, the fitness of the colourings stored in $P$ is calculated and stored in $scores$. In Lines 10 to 12, if there is a more fit individual than $\tilde{\sigma}$ in $P$, then it replaces $\tilde{\sigma}$. This procedure repeats until the termination condition is met. After the main loop ends, the algorithm reports the best colouring it has found in Line 14.

\subsection{Memetic algorithms}\label{sec:memetic}

As mentioned, MAs follow almost similar setting to GAs; the main difference is the involvement of an improvement algorithm that, at each step, enhances the solutions stored in the population. Since this improvement repeats several times during a run of the algorithm, it is very important to use a very fast and effective local search algorithm to allow time for effective evolution of the population. Based on the results of~\cite{SHEKARRIZ_local_search}, the improvement algorithm {\sf LS} is chosen due to its low time-complexity, effectiveness in such a short time, employing some randomness in its procedure, and good experimental records~\cite{SHEKARRIZ_local_search}.%

For generating the initial population of feasible colourings, again we use {\sf Rnd}, {\sf LMC}, and {\sf LS}. There was no point in improving a population generated by {\sf LS} by the same algorithm, and the result will be similar to GA's when it starts with a population of outputs of  {\sf LS}. %
Therefore, in this case, we use {\sf RLS} instead of {\sf LS} for the local search component of the algorithm. Our proposed MA is presented in Algorithm~\ref{alg:mm}.

\begin{algorithm}
	\caption{{\sf Memetic Algorithm (MA)} -- Searching for the optimal solution of $\rho$-happy colouring by combining, selecting and enhancing the solutions .}\label{alg:mm}
	\begin{algorithmic}[1]
		\State $P_0 \gets$ a population of heuristic solutions of size $PopSize$
		\State $P \gets \{Improve(G,t,\rho,V'): t\in P_0\}$
		\State $Scores\gets \{H_\rho (t), \forall t\in P\}$
		\State $\tilde{\sigma}\gets t$ for $t\in P$ such that $H_\rho (t)=\max Score$
		\NoNumber{ }
		\While{$TerminateCondition \neq$ True}
		\State $Parents\gets SelectParents(P,Scores)$
		\State $NewGen_0\gets CrossOver(Parents,PopSize, p)$\Comment{Usually $p=\frac{1}{2}$}
		\State $Mutate(NewGen_0, MuteFactor)$
		\State $NewGen \gets \{Improve(G,t,\rho,V'): t\in NewGen_0\}$
		\State $P\gets Parents \cup NewGen$
		\State $Scores\gets \{H_\rho (t), \forall t\in P\}$
		\If{$\max Score> H_\rho (\tilde{\sigma})$}
		\State $\tilde{\sigma}\gets t$ for $t\in P$ such that $H_\rho (t)=\max Score$
		\EndIf
		\EndWhile
		\State {\bf Return} $\tilde{\sigma}$
	\end{algorithmic}
\end{algorithm}

Because MAs and GAs are similar in their key evolutionary operators, we only highlight where MAs are different from GAs. %
The first is in the formation of the initial population, Lines 1 and 2 of Algorithm~\ref{alg:mm}, where an improvement algorithm, such as {\sf LS}, is run on each of the individual colourings in the initial population. The next difference is in the main loop at Line 9 of Algorithm~\ref{alg:mm}: after creating the new generation through crossover and mutation,  the improvement algorithm runs again on every single offspring stored in $NewGen_0$. The rest of the Algorithm~\ref{alg:mm} is similar to Algorithm~\ref{alg:ev}.

\section{Experimental evaluation} \label{sec:experimental}
In this section, we explain the extensive tests we run to compare our evolutionary algorithms for SHC. There are six such algorithms, namely {\sf GA(Rnd)}, {\sf GA(LMC)}, {\sf GA(LS)}, {\sf MA(Rnd)}, {\sf MA(LMC)}, and {\sf MA+RLS(LS)}. %
	For each algorithm, the name within parentheses denotes the method by which its initial populations are generated.%

First in Subsection~\ref{sec:test-dtls}, details about the instances and parameters' settings are given. We compare solutions generated by the algorithms for their number of $\rho$-happy vertices in Subsection~\ref{sec:SHC}. As said, the main objective of the algorithms is to maximise the number of $\rho$-happy vertices. Meanwhile, because a complete $\rho$-happy colouring gives a vertex partition through its colour classes, which can be considered as the graph's communities. %
	A secondary objective of utilising an algorithm to optimise an SHC in a graph that maximises $ACD(\sigma)$, which leads to a better alignment between the colour classes and the graph's community structure. We compare algorithms for this objective in Subsection~\ref{sec:ACD}. %

\subsection{Instances and parameters%
}\label{sec:test-dtls}

The EAs are tested for a set of 28,000 randomly generated partially coloured graphs in the SBM, as introduced in~\cite[Section 5.2]{SHEKARRIZ2025106893}. %
The graphs are stored in DIMACS format, the instance generator and the source code for the algorithms are publicly available\footnote{at \href{https://github.com/mhshekarriz/HappyColouring_SBM}{https://github.com/mhshekarriz/HappyColouring\_SBM}}. 

The 28,000 graphs are defined for $200 \leq n < 3,000$ vertices, 10 instances for each $n$. For each graph, other parameters are randomly chosen so that $k \in \{2,3,\ldots,20\}$, $p \in (0,1]$, $q \in (0, \frac{p}{2}]$, and $\rho \in (0,1]$. %

In each graph, a subset of the vertices is precoloured. This was done during the generation of instances, so that the colour number for each precoloured vertex agrees with its community number; if two vertices of a community are precoloured, they cannot be of different colours. %
For each graph, a random number $pcc$ is chosen from the set $\{1,2,\ldots,10\}$, and from each community, $pcc$ number of vertices are precoloured. A study on how $pcc$ can affect soft happy colouring or the accuracy of community detection of algorithms is given in~\cite[Section 6.5]{SHEKARRIZ_local_search}. 

To run an EA, we need to decide on two parameters: population size and mutation factor. Before the test, we ran the algorithms with different values of these parameters on a set of 1000 randomly generated precoloured graphs. The test shows acceptable performance of both GAs and MAs when the population size is around 20 and the mutation factor is around 0.005. We then fixed these parameters in all the tested EAs in order to have a smooth outcome comparable with each other and also with other methods.

All tests were given a time limit of 600 seconds. A lower time limit prevents the evolutionary algorithms from completing enough (at least 3) iterations of their main loops.

\subsection{Performance for the objective of SHC}\label{sec:SHC} %
\begin{figure}
\centering
\includegraphics[scale=0.45]{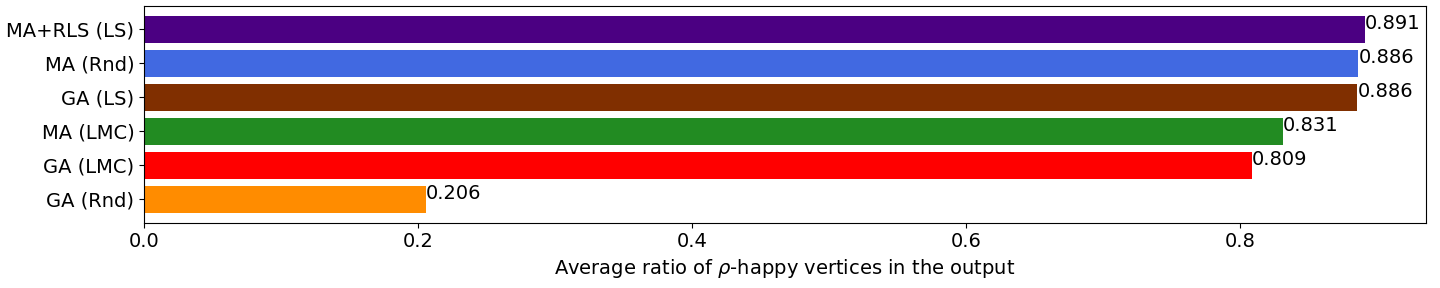}
\caption{Average ratio of $\rho$-happy vertices in the output of the evolutionary algorithms when no condition is imposed on $\rho$.%
}
\label{fig:ev-happy-bar}
\end{figure}

To compare the performances of the algorithms, a common approach is to compare their outputs in terms of the average ratio of $\rho$-happy vertices to the total number of vertices. Figure~\ref{fig:ev-happy-bar} represents this average ratio in the general case for the six tested algorithms. Based on this horizontal bar chart, on average, an output of {\sf MA+RLS(LS)} yields 89.1\% happy vertices, the best among the tested algorithms. The average ratios of $\rho$-happy vertices of {\sf MA(Rnd)} and {\sf GA(LS)} are 88.6\%, thus they are equal when there is no condition on $\rho$. Among the tested algorithms, {\sf GA(Rnd)} with 20\% has the worst average ratio of $\rho$-happy vertices, while the average performances of other algorithms are above 80\%.  

\begin{figure}
\centering
\includegraphics[scale=0.83]{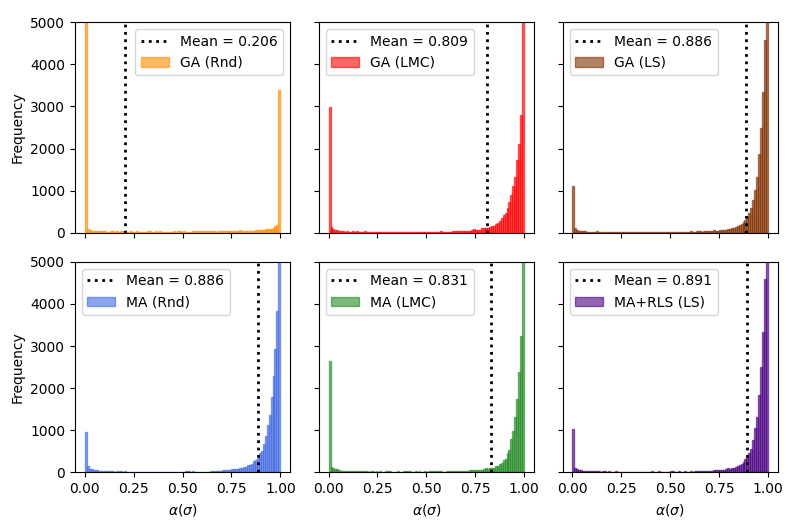}
\caption{Histogram of ratios of $\rho$-happy vertices ($\alpha (\sigma)$) of colouring outputs ($\sigma$) of the tested evolutionary algorithms. For each of the six diagrams, the dotted vertical line represents the mean value that is also reported in Figure~\ref{fig:ev-happy-bar}. The number of bins for demonstrating histograms is 100.%
}
\label{fig:Hist-ev_ave}
\end{figure}

The distribution around the average of the ratio of $\rho$-happy vertices for each of the evolutionary algorithms is shown in Figure~\ref{fig:Hist-ev_ave}. It can be seen that all the algorithms, except {\sf GA(Rnd)}, have outputs with ratios of $\rho$-happy colouring close to 1. Meanwhile, all algorithms tested experienced generating many solutions with no $\rho$-happy vertices; the least of such a failure belongs to {\sf MA(Rnd)} with fewer than 1000 outputs with no $\rho$-happy vertex, {\sf MA+RLS(LS)} and {\sf GA(LS)} come in the second and third places.  

We also tested the statistical significance of the mean values for each pair of algorithms. Table~\ref{table:stat-t-test} presents the $p$-values of each pair of algorithms based on the Welch $t$-test. %
Here, we assume $\alpha=0.05$. %
The null hypothesis is that two algorithms have no significant difference in their mean value of the ratio of $\rho$-happy vertices of their outputs. The alternative hypothesis is that these mean values are different. The null hypothesis is rejected for a pair of algorithms if the $p$-value for them is less than $\alpha$. Based on Table~\ref{table:stat-t-test}, the null hypothesis is rejected for all distinct pairs of algorithms, except the pair {\sf GA(LS)} and {\sf MA(Rnd)}. For the pair {\sf GA(LS)} and {\sf MA(Rnd)}, we can see that these two algorithms are essentially similar; in {\sf GA(LS)} the population is generated by {\sf LS}, and then the Genetic Algorithm finds the best solution it can find through recombination of the population while in {\sf MA(Rnd)} the population is randomly generated, but {\sf LS} performs on each of them before the recombination of the population happens. Therefore, both algorithms have almost similar populations of solutions after the first recombination.

\begin{table}
\small
\centering
\begin{NiceTabular}{|c|c|c|c|c|c|c|}[hvlines, code-before =
	\rectanglecolor{vibtilg!50}{1-1}{2-1}
	\rectanglecolor{C_ev!40}{3-1}{4-2}
	\rectanglecolor{C_ev!40}{1-2}{2-2}
	\rectanglecolor{C_evlmc!50}{5-1}{6-1}
	\rectanglecolor{C_evlmc!50}{5-3}{6-3}
	\rectanglecolor{C_evlmc!50}{1-3}{2-3}
	\rectanglecolor{C_evls!35}{7-1}{8-1}
	\rectanglecolor{C_evls!35}{7-4}{8-4}
	\rectanglecolor{C_evls!35}{1-4}{2-4}
	\rectanglecolor{C_mm!50}{9-1}{10-1}
	\rectanglecolor{C_mm!50}{9-5}{10-5}
	\rectanglecolor{C_mm!50}{1-5}{2-5}
	\rectanglecolor{C_mmlmc!40}{11-1}{12-1}
	\rectanglecolor{C_mmlmc!40}{11-6}{12-6}
	\rectanglecolor{C_mmlmc!40}{1-6}{2-6}
	\rectanglecolor{C_mmrls!40}{13-1}{14-1}
	\rectanglecolor{C_mmrls!40}{13-7}{14-7}
	\rectanglecolor{C_mmrls!40}{1-7}{2-7}
	]
	\hline

	\Block{2-1}{ Algorithms} & \Block{2-1}{ \sf GA \\ (Rnd)} & \Block{2-1}{\sf GA \\ (LMC)} & \Block{2-1}{\sf GA \\ (LS)}
	& \Block{2-1}{\sf MA \\ (Rnd)} & \Block{2-1}{\sf MA \\(LMC)} & \Block{2-1}{ \sf MA+RLS \\ (LS) }\\ &&&&&&\\
	\hline
	\Block{2-1}{\sf GA(Rnd)}
	& \Block{2-1}{1}           & \Block{2-1}{0}            & \Block{2-1}{0}
	& \Block{2-1}{0}            & \Block{2-1}{0}            & \Block{2-1}{0}             \\&&&&&&\\
	\hline
	\Block{2-1}{\sf GA(LMC)}
	&\Block{2-1}{0}           & \Block{2-1}{1}            & \Block{2-1}{0} 
	& \Block{2-1}{0}            & \Block{2-1}{0}            & \Block{2-1}{0}             \\&&&&&&\\
	\hline
	\Block{2-1}{\sf GA(LS)}
	& \Block{2-1}{0}           & \Block{2-1}{0}            & \Block{2-1}{1}
	& \Block{2-1}{\textcolor{red}{0.760}}
	& \Block{2-1}{0}            & \Block{2-1}{0.009}            \\&&&&&&\\
	\hline
	\Block{2-1}{\sf MA(Rnd)}
	& \Block{2-1}{0}           & \Block{2-1}{0}            & \Block{2-1}{\textcolor{red}{0.760}}
	& \Block{2-1}{1}            & \Block{2-1}{0}            & \Block{2-1}{0.022}            \\&&&&&&\\
	\hline
	\Block{2-1}{\sf MA(LMC)}
	& \Block{2-1}{0}           & \Block{2-1}{0}            & \Block{2-1}{0}
	& \Block{2-1}{0}            & \Block{2-1}{1}            & \Block{2-1}{0}             \\&&&&&&\\
	\hline
	\Block{2-1}{\sf MA+RLS \\(LS)}
	& \Block{2-1}{0}           & \Block{2-1}{0}            & \Block{2-1}{0.009}
	& \Block{2-1}{0.022}       & \Block{2-1}{0}            & \Block{2-1}{1}             \\&&&&&&\\
	\hline
\end{NiceTabular}

\medskip
\caption{Statistical $p$-values of Welch’s $t$-test for each pair of tested evolutionary algorithms. The null hypothesis is that there is no significant difference between the mean values of two algorithms. This hypothesis is rejected if the $p$-value is greater than 0.05. It is rejected for all pairs of distinct algorithms but for the pair {\sf GA(LS)} and {\sf MA(Rnd)}. To simplify, when a $p$-value is less than $1\times 10^{-16}$, we have put 0 in the table.
}
\label{table:stat-t-test}
\end{table}

\begin{figure}
\centering
\includegraphics[scale=0.45]{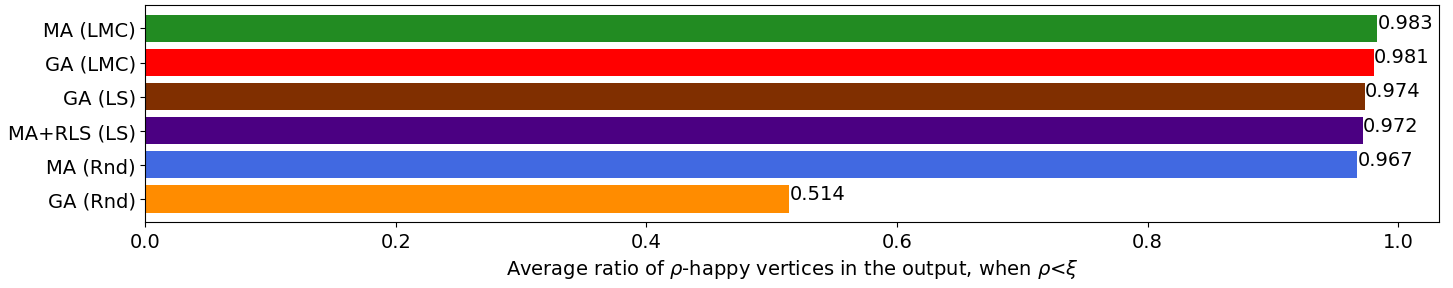}
\caption{Average ratio of $\rho$-happy vertices in the output of the evolutionary algorithms when $0\le\rho\le \xi$.}
\label{fig:ev-happy-bar-xi}
\end{figure}

Based on the theoretical work in~\cite[Theorem 3.2]{SHEKARRIZ2025106893}, we expect to have a higher ratio of $\rho$-happy vertices when $\rho<\xi$. Figure~\ref{fig:ev-happy-bar-xi} compares this average ratio for the tested algorithms when the condition $\rho<\xi$ is met. In this case, the algorithms whose populations are generated by {\sf LMC} illustrate a very high average ratio of $\rho$-happy vertices. On average, the outputs of {\sf MA(LMC)} and {\sf GA(LMC)} have respectively 98.3\% and 98.1\% of vertices $\rho$-happy, earning them the first and the second place when $\rho<\xi$. In this case, the average performance of {\sf GA(Rnd)} toward the objective of SHC reaches 51\%, far below other algorithms whose average ratios of $\rho$-happy vertices are at least 96\%. These average values and their standard deviations are reported in Table~\ref{table:ev-hap-SD}. 

\begin{table}
\centering
\begin{NiceTabular}{|c|cc|cc|cc|}[hvlines, code-before =
	\rectanglecolor{vibtilg!50}{1-1}{1-7}
	\rectanglecolor{des!50}{2-1}{2-7}
	\rectanglecolor{C_ev!40}{3-1}{3-7}
	\rectanglecolor{C_evlmc!50}{4-1}{4-7}
	\rectanglecolor{C_evls!35}{5-1}{5-7}
	\rectanglecolor{C_mm!50}{6-1}{6-7}
	\rectanglecolor{C_mmlmc!40}{7-1}{7-7}
	\rectanglecolor{C_mmrls!40}{8-1}{8-7}
	]
	\hline
	{\bf Condition} & \multicolumn{2}{c|}{$\boldsymbol{0\leq \rho\leq 1}$} & \multicolumn{2}{c|}{$\boldsymbol{0\leq \rho\leq\xi}$} & \multicolumn{2}{c|}{$\boldsymbol{\xi< \rho\leq 1}$} \\
	\hline
	Description  & {\bf mean}  & {\bf $SD$}  & {\bf mean}  & {\bf $SD$}   & {\bf mean}  & {\bf $SD$}    \\
	\hline
	{\sf GA(Rnd)}  & 0.206  & 0.378  & 0.514  & 0.451  & 0.033  & {\bf 0.162}   \\
	\hline
	{\sf GA(LMC)} & 0.809  & 0.335  & 0.981  & 0.047  & 0.712  & 0.384  \\
	\hline
	{\sf GA(LS)}  & 0.886  & 0.241  & 0.974  & {\bf 0.036}  & 0.837  & 0.288   \\
	\hline
	{\sf MA(Rnd)} & 0.886  & 0.239  & 0.967  & 0.068  & 0.841  & 0.284  \\
	\hline
	{\sf MA(LMC)}  & 0.831  & 0.324  & {\bf 0.983}  & 0.058  & 0.746  & 0.376   \\
	\hline
	{\sf MA+RLS(LS)} & {\bf 0.891}  & {\bf 0.235}  & 0.972  & 0.064  & {\bf 0.846}  & 0.280   \\
	\hline
\end{NiceTabular}

\medskip
\caption{Mean values and standard deviation of ratios of $\rho$-happy vertices, when no condition imposed on $\rho$, when $\rho\in[0,\xi]$, and when $\rho\in (\xi, 1]$.}
\label{table:ev-hap-SD}
\end{table}

\begin{figure}
\centering
\includegraphics[scale=0.45]{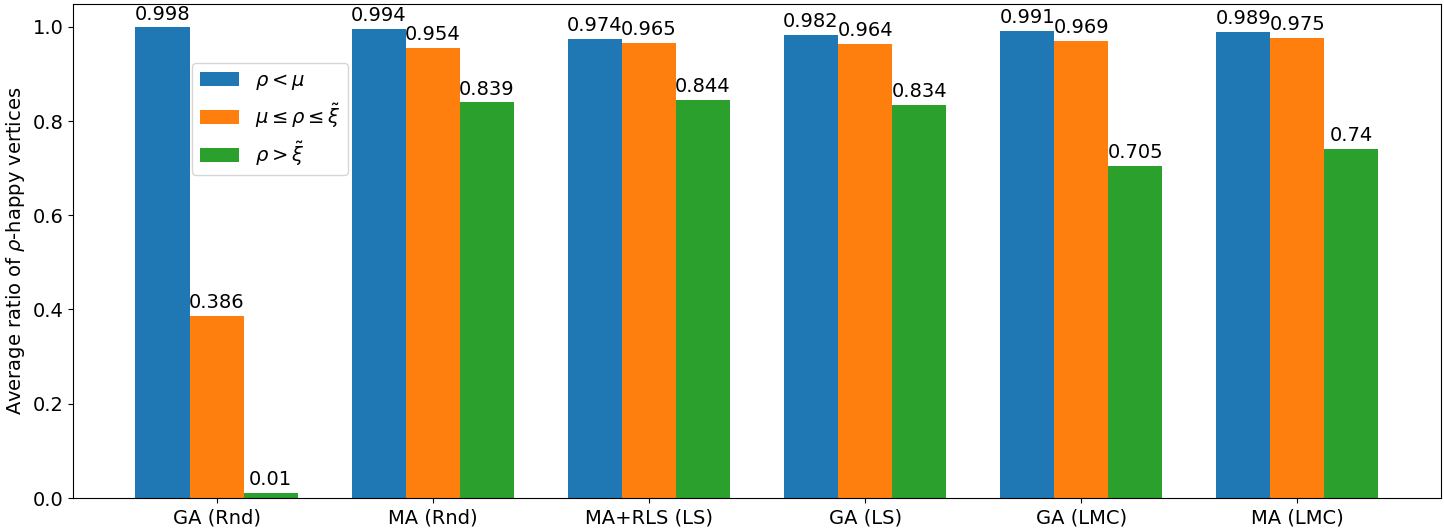}
\caption{Average ratio of $\rho$-happy vertices in the output of the evolutionary algorithms when $\rho<\mu$, $\mu\le\rho\le \tilde{\xi}$, or $\rho >\tilde{\xi}$.}
\label{fig:ev-happy-bar-xid-xit}
\end{figure}

Theoretical results~\cite[Theorems 3.1 and 3.2]{SHEKARRIZ_local_search} also say that the average performance of the algorithms for $\rho<\mu$, $\mu\le\rho\le \tilde{\xi}$, or $\rho >\tilde{\xi}$ are radically different. Figure~\ref{fig:ev-happy-bar-xid-xit} provides a comparison of the average ratios of $\rho$-happy vertices of outputs of the six evolutionary algorithms tested for each of these cases. 

For the case $\rho<\mu$, all six algorithms perform very well, see blue columns in Figure~\ref{fig:ev-happy-bar-xid-xit}. The worst among them is {\sf MA+RLS(LS)} with an average of 97.4\% for the ratio of $\rho$-happy vertices, and the best is {\sf GA(Rnd)} with an average of 99.8\%. These high averages are not unexpected because the lower $\rho$ leads SHC to an easier optimisation problem.

The case $\mu\le\rho\le \tilde{\xi}$ is important because the highest expected values of community detection of algorithms happen in this case~\cite{SHEKARRIZ_local_search}. Average ratios of $\rho$-happy vertices are demonstrated by orange bars in Figure~\ref{fig:ev-happy-bar-xid-xit}. As expected, these average ratios are slightly lower than in case $\rho<\mu$, except for {\sf GA(Rnd)} which shows a sharp drop. The best performance here belongs to {\sf MA(LMC)} with an average of 97.5\% of $\rho$-happy vertices.

The case $\rho>\tilde{\xi}$ is the most difficult case because finding a complete $\rho$-happy colouring is theoretically impossible for a graph in the SBM with an ample number of vertices~\cite[Theorem 3.2]{SHEKARRIZ_local_search}. The green bars in Figure~\ref{fig:ev-happy-bar-xid-xit} give the average values for the ratio of $\rho$-happy vertices in the solutions tested by the EAs for this case. Here, {\sf MA+RLS(LS)} has the best average of 84.4\% while {\sf GA(Rnd)} is the worst with an average of 1\%.

The number of times that each algorithm can produce a complete $\rho$-happy colouring can be another criterion when comparing them. For example, for the same set of 28,000 graphs we are testing the six EAs, {\sf LMC} could find 1,876 complete $\rho$-happy colourings, 635 of which for $0\le \rho<\mu$, 1,241 when $\mu\le \rho\le \tilde{\xi}$, and no complete $\rho$-happy colouring when $\rho>\tilde{\xi}$~\cite[Table 1]{SHEKARRIZ_local_search}. Table~\ref{table:ev-hap} gives similar information about the EAs tested. The most notable items of this table are for {\sf MA(LMC)} that has 3,014 complete $\rho$-happy colourings when $\mu\le \rho\le \tilde{\xi}$ and 1,280 ones when $0\le \rho<\mu$, which makes it a total of 4,294 complete $\rho$-happy colourings; more than 2.2 times of {\sf LMC}'s. While {\sf MA(LMC)} seems to be unbeatable for the total number of complete $\rho$-happy colourings and also for the case $\mu\le \rho\le \tilde{\xi}$, {\sf GA(LMC)} holds the second place with 3644 complete $\rho$-happy colourings in total. For the case $0\le \rho<\mu$, {\sf GA(Rnd)} gives the best performance of finding complete $\rho$-happy colourings, though this case is not as important as the case $\mu\le \rho\le \tilde{\xi}$. The worst algorithms in this context are {\sf GA(LS)} and {\sf MA+RLS(LS)}, whose total number of $\rho$-happy colourings are respectively 580 and 432 out of 28,000. 

\begin{table}
\centering
\begin{NiceTabular}{|c|cc|cc|cc|}[hvlines, code-before =
	\rectanglecolor{vibtilg!50}{1-1}{1-7}
	\rectanglecolor{des!50}{2-1}{2-7}
	\rectanglecolor{C_ev!40}{3-1}{3-7}
	\rectanglecolor{C_evlmc!50}{4-1}{4-7}
	\rectanglecolor{C_evls!35}{5-1}{5-7}
	\rectanglecolor{C_mm!50}{6-1}{6-7}
	\rectanglecolor{C_mmlmc!40}{7-1}{7-7}
	\rectanglecolor{C_mmrls!40}{8-1}{8-7}
	\rectanglecolor{des!50}{9-1}{10-7}
	]
	\hline
	{\bf Condition} & \multicolumn{2}{c|}{$\boldsymbol{0\leq \rho<\mu}$} & \multicolumn{2}{c|}{$\boldsymbol{\mu\leq \rho\leq\tilde{\xi}}$} & \multicolumn{2}{c|}{$\boldsymbol{\tilde{\xi}< \rho\leq 1}$} \\
	\hline
	Description  & \# $\sigma\in H_\rho$  & \# $\sigma\notin H_\rho$  & \# $\sigma\in H_\rho$  & \# $\sigma\notin H_\rho$  & \# $\sigma\in H_\rho$    & \# $\sigma\notin H_\rho$   \\
	\hline
	{\sf GA(Rnd)}  & {\bf 2151}  & {\bf 201}  & 946  & 7461  & 0  & 17241   \\
	\hline
	{\sf GA(LMC)} & 1368  & 984  & 2276  & 6131  & 0  & 17241  \\
	\hline
	{\sf GA(LS)}  & 424  & 1928  & 156  & 8251  & 0  & 17241   \\
	\hline
	{\sf MA(Rnd)} & 2051  & 301  & 953  & 7454  & 0  & 17241  \\
	\hline
	{\sf MA(LMC)}  & 1280  & 1072  & {\bf 3014}  & {\bf 5393}  & 0  & 17241   \\
	\hline
	{\sf MA+RLS(LS)} & 236  & 2116  & 196  & 8211  & 0  & 17241   \\
	\hline
	\Block{2-1}{Totals} &  \multicolumn{2}{c|}{2352} &  \multicolumn{2}{c|}{8407} &  \multicolumn{2}{c|}{17241}\\
	&  \multicolumn{6}{c|}{28000}\\ 
	\hline
\end{NiceTabular}

\medskip
\caption{The number of times each of six evolutionary algorithms finds a complete $\rho$-happy colouring (denoted by \# $\sigma\in H_\rho$) versus the number of times each of them fails to find a complete $\rho$-happy colouring (denoted by \# $\sigma\notin H_\rho$). There were a total of 28,000 graphs in the test. The numbers are reported for cases  $\rho \in [0,\mu)$, $\rho \in [\mu, \tilde{\xi}]$, and $\rho \in (\tilde{\xi},1]$. When $\rho\in (\tilde{\xi},1]$, none of the algorithms could find a complete $\rho$-happy colouring, so the respective value is 0.}
\label{table:ev-hap}
\end{table}

\begin{figure}
\captionsetup{size=small}
\begin{subfigure}{0.5\textwidth}
	\includegraphics[scale=0.48]{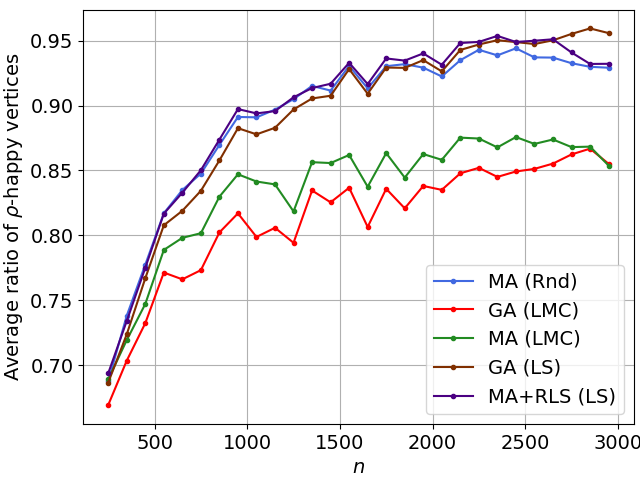}
	\caption{}\label{fig:ev-hap-n}
\end{subfigure} 
\begin{subfigure}{0.5\textwidth}
	\includegraphics[scale=0.48]{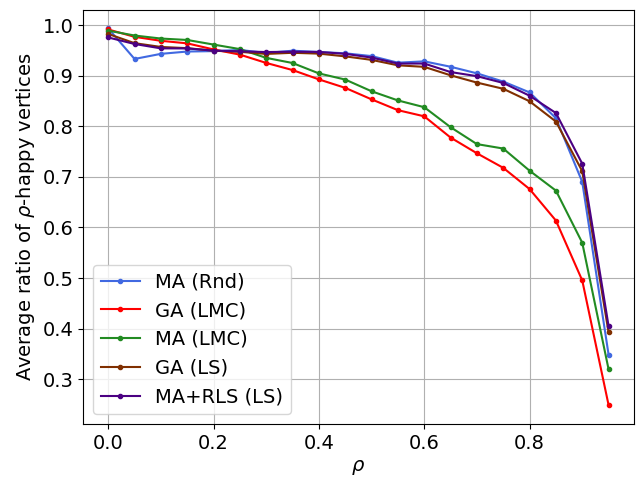}
	\caption{}\label{fig:ev-hap-r}
\end{subfigure} 

\begin{subfigure}{1\textwidth}
	\centering
	\includegraphics[scale=0.5]{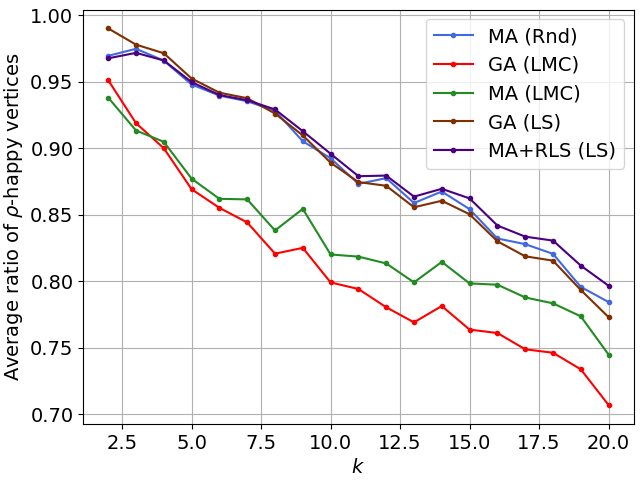}
	\caption{}\label{fig:ev-hap-k}
\end{subfigure} 

\caption{Comparison of the GAs and MAs for their average number of $\rho$-happy vertices considering (a) the number of vertices $n$, (b) the proportion of happiness $\rho$, and (c) the number of colours $k$. For visualisation of these charts, the algorithm {\sf GA(Rnd)} is excluded because its poor performance is not comparable with other algorithms and adding it to the charts makes the performance of other algorithms almost indistinguishable.}\label{fig:ev-hap}
\end{figure}

For each algorithm, the average ratio of $\rho$-happy vertices to the number of vertices changes with other parameters. It is especially interesting to see its trends regarding changes in the number of vertices $n$, the proportion of happiness $\rho$, and the number of colours. Figure~\ref{fig:ev-hap} presents these trends for five of the six evolutionary algorithms tested; {\sf GA(Rnd)} is excluded from this figure, because its poor performance illustrated in the figures, makes the other five indistinguishable.

As expected by the theoretical results in~\cite{SHEKARRIZ2025106893}, increasing the number of vertices leads to higher ratios of $\rho$-happy vertices in a $k$-colouring of a graph in the SBM when $k$ remains unchanged. Our experimental results for the evolutionary algorithms verify this; see Figure~\ref{fig:ev-hap-n}. We see that the ratio of $\rho$-happy vertices of outputs by the algorithms whose initial populations are made by {\sf LMC}, i.e. {\sf GA(LMC)} and {\sf MA(LMC)}, cannot reach 90\%, while algorithms using {\sf LS}, i.e. {\sf GA(LS)}, {\sf MA+RLS(LS)}, and interestingly {\sf MA(Rnd)} have ratios of $\rho$-happy vertices above 90\% when the number of vertices is greater than 1500 and $k\le 20$. The algorithm {\sf GA(LS)} is outperformed by {\sf MA(Rnd)} and {\sf MA+RLS(LS)} when $n<1800$, but afterwards reaches the first place, especially when $n>2600$.

In Figure~\ref{fig:ev-hap-r}, average ratios of $\rho$-happy vertices are illustrated with changes in the proportion of happiness $\rho$. The falling trend can be seen in this ratio for each of the EAs. However, when $\rho<0.2$, algorithms with an initial population made by {\sf LMC} tend to generate more $\rho$-happy vertices. This observation justifies the high average ratio of $\rho$-happy vertices in {\sf MA(LMC)} and {\sf GA(LMC)} in Figure~\ref{fig:ev-happy-bar-xi}, by noting that in our tests the average value for $\xi$ is around 0.35.

Figure~\ref{fig:ev-hap-k} demonstrates the performance of evolutionary algorithms, all but {\sf GA(Rnd)}, concerning changes in the number of colours. It is evident from the figure that increasing the number of colours negatively affects the average ratio of $\rho$-happy vertices that can be found by an algorithm. This average ratio for {\sf MA+RLS(LS)}, {\sf MA(Rnd)} and {\sf GA(LS)} is much higher than that of {\sf MA(LMC)} and {\sf GA(LMC)}. Meanwhile, when $2\leq k<6$, {\sf GA(LS)} produces colourings with the highest average ratio of $\rho$-happy vertices. When $8<k\leq 20$, it is {\sf MA+RLS(LS)} that gives the best solution on average. The average ratio of $\rho$-happy vertices of solutions created by {\sf MA(Rnd)} lies between that of {\sf MA+RLS(LS)} and {\sf GA(LS)}. Interestingly, when $6\leq k\leq 8$, all three algorithms generate solutions with almost equal average ratios of $\rho$-happy vertices.

\subsection{Accuracy of community detection}\label{sec:ACD}

\begin{figure}
\centering
\includegraphics[scale=0.45]{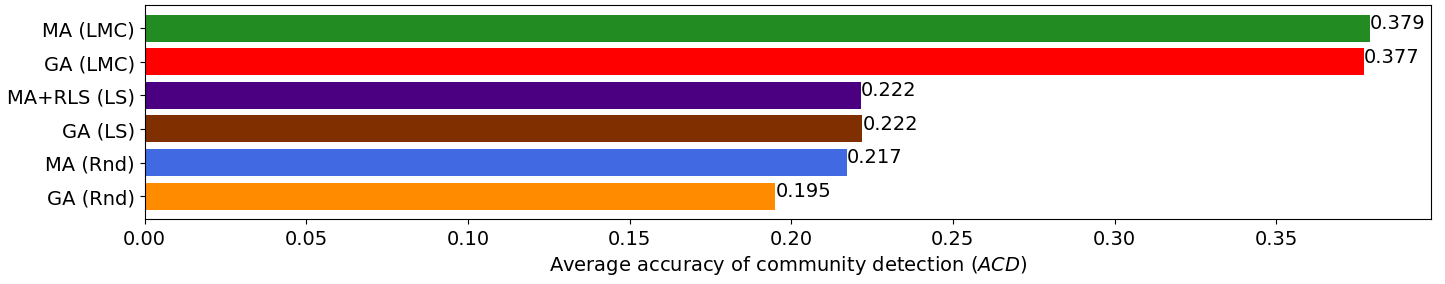}
\caption{Average accuracy of community detection in the output of the evolutionary algorithms when no condition is imposed on $\rho$.}
\label{fig:ev-comm}
\end{figure}

Another important factor in comparing the quality of a solution $\sigma$ is the accuracy of community detection, $ACD(\sigma)$. %
To compare the six tested EAs, let us start by comparing the average values of $ACD(\sigma)$ of the outputs of these algorithms by looking at Figure~\ref{fig:ev-comm}. From the figure, it is obvious that {\sf MA(LMC)} and {\sf GA(LMC)} have the highest average accuracy of community detection by far of over 37\%. Other algorithms' average accuracy of community detection is around 20\%. Moreover, {\sf LMC} had an average 46.2\% of accuracy of community detection (see\cite[Figure 7]{SHEKARRIZ_local_search}. Therefore, in the general case, the EAs do not necessarily push the solutions towards more aligned colour classes with original communities. However, as we will soon see, the situation when $\mu\le\rho\le\tilde{\xi}$ is entirely different. %

\begin{figure}
\centering
\includegraphics[scale=0.46]{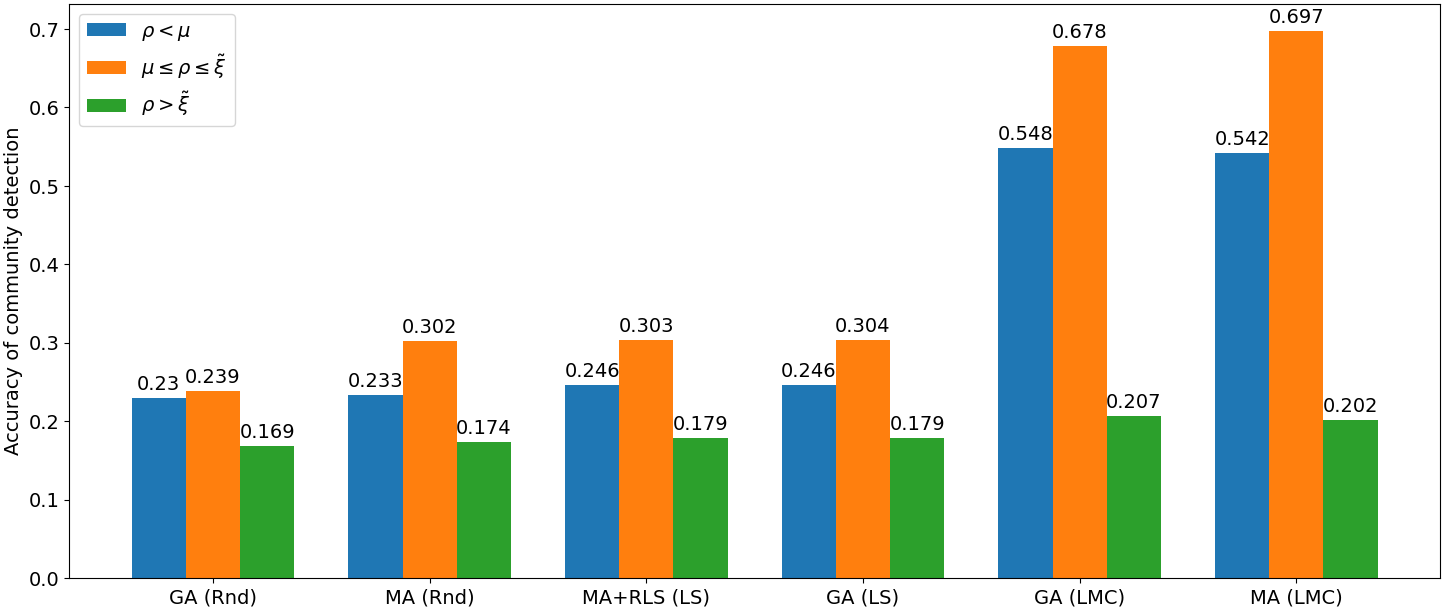}
\caption{Average accuracy community detection ($ACD$) of the tested evolutionary algorithms.}
\label{fig:th2}
\end{figure}

The best among known heuristics in terms of accuracy of community detection is {\sf LMC}, with an average of 46.2\% in general and an average of 64\% when $\mu\le\rho\le\tilde{\xi}$, %
for the set of graphs in the SBM we tested, see~\cite[Figure 15]{SHEKARRIZ_local_search}. However, as can be seen through orange bars in Figure~\ref{fig:th2}, {\sf MA(LMC)} improved this average by more than 5\%, making it 69.7\% when $\mu\le\rho\le\tilde{\xi}$. The algorithm {\sf GA(LMC)} with the accuracy of 67.8\% earns the second-best place in this case. For the case $\rho<\mu$, again the two algorithms with initial population generated by {\sf LMC} have the highest accuracy of community detection; {\sf MA(LMC)} and {\sf GA(LMC)} improved the accuracy of community detection of {\sf LMC} from 46.5\%~\cite[Figure~4]{SHEKARRIZ_local_search} to respectively 54.2\% and 54.8\%. For the case $\rho>\tilde{\xi}$, all six tested evolutionary algorithms have accuracies of community detection between 16.9\% to 20.7\%.

\begin{figure}
\centering
\includegraphics[scale=0.46]{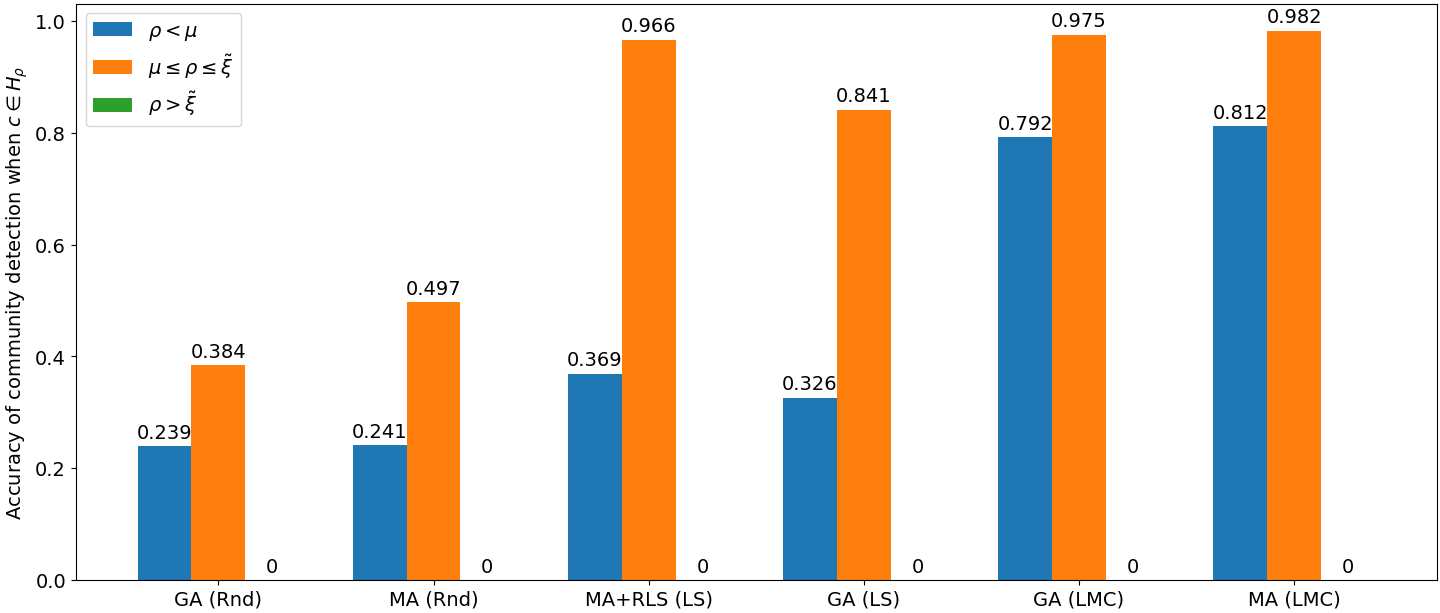}
\caption{Average accuracy community detection ($ACD$) when the evolutionary algorithms found complete $\rho$-happy colourings. When $\rho >\tilde{\xi}$, no algorithm could find a complete $\rho$-happy colouring. Consequently, no green bars can be seen on the chart.}
\label{fig:th1}
\end{figure}

When the six EAs %
generate a complete $\rho$-happy solution $\sigma$, the accuracy of community detection is expected to be the highest possible. Figure~\ref{fig:th1} visualised the average $ACD(\sigma)$ when $\rho<\mu$, $\mu\leq \rho\leq\tilde{\xi}$ and $\rho>\tilde{\xi}$. For the case $\rho<\mu$, the blue bars of Figure~\ref{fig:th1} show that the highest average $ACD(\sigma)$ belongs to {\sf MA(LMC)} with 81.2\%, followed by {\sf GA(LMC)} with 79.2\%, while other algorithms demonstrate unimpressive average $ACD(\sigma)$ in this case. 

However, when $\mu\leq \rho\leq\tilde{\xi}$, the average $ACD(\sigma)$ for a complete $\rho$-happy colouring $\sigma$ reached impressive average values of 98.2\% for {\sf MA(LMC)}, 97.5\% for {\sf GA(LMC)}, and 96.6\% for {\sf MA+RLS(LS)}. When $\rho>\tilde{\xi}$, none of the six tested evolutionary algorithms could find a complete $\rho$-happy colouring, hence no green bar can be seen in Figure~\ref{fig:th1}. Indeed, this has been previously predicted for soft happy colouring algorithms ~\cite[Theorem 3.2]{SHEKARRIZ_local_search}.

We can also compare algorithms using the number of times colour classes of the colouring output of each algorithm accurately align with the original communities of the graphs, whether or not they are complete $\rho$-happy colourings. %
Table~\ref{table:ev-comm} presents the results for the three cases $0\leq \rho <\mu$, $\mu\leq \rho\leq\tilde{\xi}$ and $\rho>\tilde{\xi}$ (and the respective number of times each algorithm fails to detect communities accurately).

\begin{table}
\centering
\begin{NiceTabular}{|c|cc|cc|cc|}[hvlines, code-before =
	\rectanglecolor{vibtilg!50}{1-1}{1-7}
	\rectanglecolor{des!50}{2-1}{2-7}
	\rectanglecolor{C_ev!40}{3-1}{3-7}
	\rectanglecolor{C_evlmc!50}{4-1}{4-7}
	\rectanglecolor{C_evls!35}{5-1}{5-7}
	\rectanglecolor{C_mm!50}{6-1}{6-7}
	\rectanglecolor{C_mmlmc!40}{7-1}{7-7}
	\rectanglecolor{C_mmrls!40}{8-1}{8-7}
	\rectanglecolor{des!50}{9-1}{10-7}
	]
	\hline
	{\bf Condition} & \multicolumn{2}{c|}{$\boldsymbol{0\leq \rho<\mu}$} & \multicolumn{2}{c|}{$\boldsymbol{\mu\leq \rho\leq\tilde{\xi}}$} & \multicolumn{2}{c|}{$\boldsymbol{\tilde{\xi}< \rho\leq 1}$} \\
	\hline
	Description  & $ACD$  & $\cancel{ACD}$  &  $ACD$  & $\cancel{ACD}$  &  $ACD$  & $\cancel{ACD}$   \\
	\hline
	{\sf GA(Rnd)}  & 0  & 2352  & 5  & 8402  & 0  & 17241   \\
	\hline
	{\sf GA(LMC)} & {\bf 9}  & 2343  & 789  & 7618  & 24  & 17217  \\
	\hline
	{\sf GA(LS)}  & 0  & 2352  & 52  & 8355  & 1  & 17240   \\
	\hline
	{\sf MA(Rnd)} & 0  & 2352  & 71  & 8336  & 0  & 17241  \\
	\hline
	{\sf MA(LMC)}  & 2  & 2350  & {\bf 1000}  & 7407  & {\bf 44}  & 17197   \\
	\hline
	{\sf MA+RLS(LS)} & 0  & 2352  & 84  & 8323  & 10  & 17231   \\
	\hline
	\Block{2-1}{Totals} &  \multicolumn{2}{c|}{2352} &  \multicolumn{2}{c|}{8407} &  \multicolumn{2}{c|}{17241}\\
	&  \multicolumn{6}{c|}{28000}\\ 
	\hline
\end{NiceTabular}

\medskip
\caption{The number of times each of six evolutionary algorithms accurately detects communities of the tested graphs (denoted by $ACD$) versus the number of times each of them fails to do so (denoted by $\cancel{ACD}$). The total number of tested graphs was 28,000. The numbers reported for cases  $\rho \in [0,\mu)$, $\rho \in [\mu, \tilde{\xi}]$, and $\rho \in (\tilde{\xi},1]$.}
\label{table:ev-comm}
\end{table}

Table~\ref{table:ev-comm} shows that {\sf MA(LMC)} has accurately found communities of 1,046 graphs out of 28,000, which is the best among the tested algorithms.  The second-best algorithm of this kind is {\sf GA(LMC)}, which has accurately found communities of 822 graphs. Interestingly, both these algorithms can accurately find communities of graphs even when $\rho>\tilde{\xi}$ for which finding a complete $\rho$-happy colouring was not possible. In this case, {\sf MA(LMC)} found 44 and {\sf GA(LMC)} found 24 colourings accurately aligned with the communities of the graphs. Because of the theoretical results of~\cite{SHEKARRIZ_local_search}, it is not unexpected that algorithms could not accurately find communities of large numbers of graphs when $0\leq \rho<\mu$.  

\begin{figure}
\captionsetup{size=small}
\begin{subfigure}{0.5\textwidth}
	\includegraphics[scale=0.5]{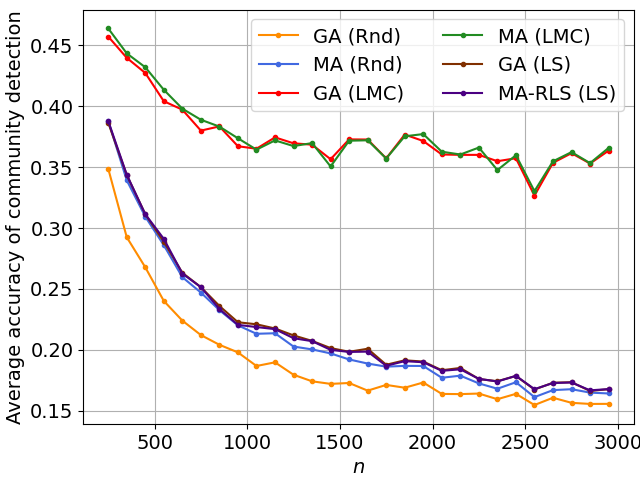}
	\caption{}\label{fig:ev-comm-n}
\end{subfigure} 
\begin{subfigure}{0.5\textwidth}
	\includegraphics[scale=0.5]{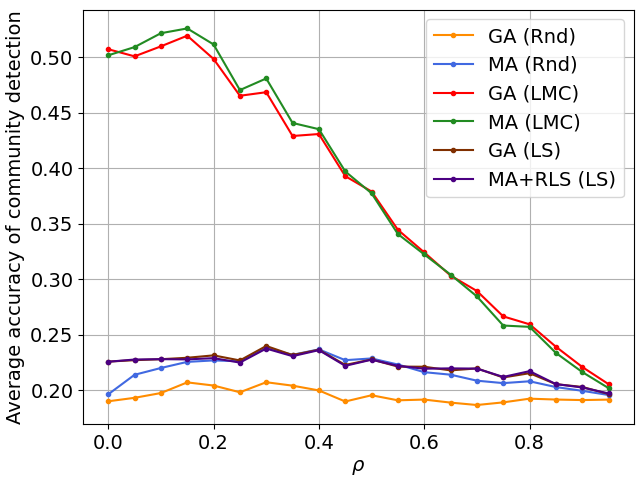}
	\caption{}\label{fig:ev-comm-r}
\end{subfigure} 

\begin{subfigure}{1\textwidth}
	\centering
	\includegraphics[scale=0.5]{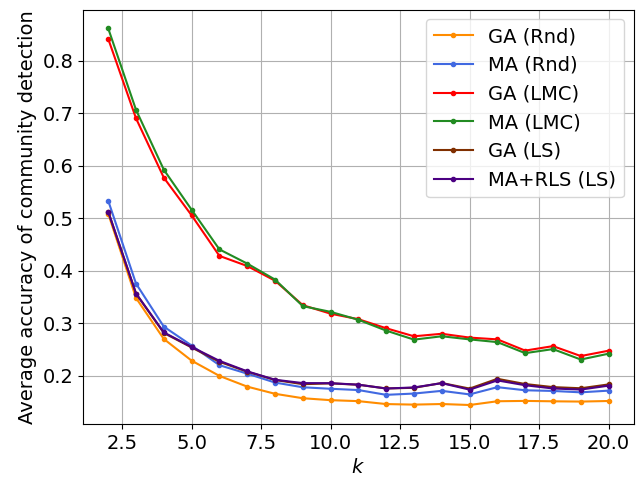}
	\caption{}\label{fig:ev-comm-k}
\end{subfigure} 

\caption{Comparison of the GAs and MAs for their accuracy of community detection considering (a) the number of vertices $n$ (b) the proportion of happiness $\rho$ and (c) the number of colours $k$.}\label{fig:ev-comm-nrk}
\end{figure}

Figure~\ref{fig:ev-comm-nrk} articulates the changes in the accuracy of community detection when the number of vertices $n$, the proportion of happiness $\rho$, or the number of colours $k$ changes. It shows that {\sf MA(LMC)} and {\sf GA(LMC)} have the greatest average accuracy of community detection. While other algorithms' accuracies of community detection tend to 0 when the number of vertices increases, these two algorithms show that the average values of $ACD(\sigma)$ remain above 0.33. Figure~\ref{fig:ev-comm-r} reveals that the tested EAs, except those whose initial populations are generated by {\sf LMC}, have low accuracies of community detection in general, which are minimally influenced by changes in $\rho$. It is not surprising that when $\rho$ gets very close to 1, the accuracy of community detections of these algorithms tends to 0 because, in this case, finding $\rho$-happy vertices is usually difficult. From Figure~\ref{fig:ev-comm-k}, the negative effect of increasing the number of colours $k$ is evident on the average accuracy of community detection of each of the tested algorithms.

\section{Conclusion}\label{sec:conc}

This study successfully developed and evaluated Memetic Algorithms (MAs) for the Soft Happy Colouring (SHC) problem, establishing a new benchmark for performance on Stochastic Block Models. By conducting a comprehensive comparison of evolutionary frameworks initialised with {\sf Local Maximal Colouring} ({\sf LMC}), {\sf Local Search} ({\sf LS}), and random inputs, we identified critical factors for maximising $\rho$-happy vertices and recovering community structures.

Our experimental results demonstrate the clear superiority of the hybrid approach. Specifically, {\sf MA+RLS(LS)} dominated the test field, achieving the highest average ratio of $\rho$-happy vertices. Furthermore, {\sf MA(LMC)} proved to be robust in specific topological conditions (where $\mu\leq \rho\leq \tilde{\xi}$), yielding the highest average accuracy for community detection. The comparative analysis highlights the vital importance of the improvement algorithm: the drastic performance gap between {\sf MA(Rnd)} and {\sf GA(Rnd)} confirms that the integration of {\sf LS} is not merely beneficial but essential for escaping local optima in this domain.

These findings empirically validate the theoretical propositions of~\cite{SHEKARRIZ_local_search, SHEKARRIZ2025106893} and underscore the efficiency of {\sf LS} as a powerful operator. Given its ability to deliver substantial improvements within short runtimes, the {\sf LS} component offers a promising foundation for designing future metaheuristics and matheuristics beyond standard evolutionary computation. Finally, while these heuristic approaches have proven effective, future research should also explore exact approaches to further define the complexity boundaries of SHC.

\section*{Declarations of interest} 
None

\bibliographystyle{plain}
\bibliography{HC.bib}

\end{document}